\documentclass[%
 reprint,
nobibnotes,
 amsmath,amssymb,
 aps,
]{revtex4-2}

\usepackage{graphicx}% Include figure files
\usepackage{dcolumn}% Align table columns on decimal point
\usepackage{bm}% bold math
\usepackage{mathtools}
\usepackage{braket}
\usepackage[caption=false, position=top]{subfig}

\usepackage{hyperref}% add hypertext capabilities
\newcolumntype{P}[1]{>{\centering\arraybackslash}p{#1}}
\newcolumntype{M}[1]{>{\centering\arraybackslash}m{#1}}

\newcommand{\eps}{\varepsilon}
\renewcommand{\vec}[1]{\boldsymbol{\mathrm{#1}}}
\renewcommand{\exp}[1]{\text{e}^{#1}}

\newcommand{\op}[1]{\hat{#1}}
\newcommand{\vop}[1]{\boldsymbol{\hat{\mathrm{#1}}}}

\newcommand{\unvec}[1]{\boldsymbol{\hat{\mathrm{#1}}}}
\newcommand{\abs}[1]{|#1|}

\newcommand{\MoS}{MoS\textsubscript{2}}
\newcommand{\MoSe}{MoSe\textsubscript{2}}
\newcommand{\WS}{WS\textsubscript{2}}
\newcommand{\WSe}{WSe\textsubscript{2}}

\begin{document}

\preprint{APS/123-QED}

\title{Optical Emission from Light-like and Particle-like Excitons in Monolayer Transition Metal Dichalcogenides}

\author{Mikkel Ohm Sauer$^{1,2}$}
\email{mikkelos@mp.aau.dk}
\author{Carl Emil Mørch Nielsen$^{1}$}
\author{Lars Merring-Mikkelsen$^{1}$}
\author{Thomas Garm Pedersen$^{1,2}$}
\email{tgp@mp.aau.dk}
\affiliation{$^1$Department of Materials and Production, Aalborg University, 9220 Aalborg {\O}st, Denmark}
\affiliation{$^2$Center for Nanostructured Graphene (CNG), 9220 Aalborg {\O}st, Denmark}

\date{\today}% It is always \today, today,

\begin{abstract}
Several monolayer transition metal dichalcogenides (TMDs) are direct band gap semiconductors and potentially efficient emitters in light emitting devices. Photons are emitted when strongly bound excitons decay radiatively, and accurate models of such excitons are important for a full understanding of the emission. Importantly, photons are emitted in directions uniquely determined by the exciton center of mass momentum and with lifetimes determined by the exciton transition matrix element. The exciton band structures of two-dimensional hexagonal materials, including TMDs, are highly unusual with coexisting particle- and light-like bands. The latter is non-analytic with emission selection rules essentially opposite to the particle-like states, but has been ignored in analyses of TMD light emission so far. In the present work, we analyse the temperature and angular dependence of light emission from both exciton species and point out several important consequences of the unique exciton band structure. Within a first-principles Density-Functional-Theory+Bethe-Salpeter-Equation framework, we compute exciton band structures and optical matrix elements for the important TMDs  {\MoS}, {\MoSe}, {\WS}, and {\WSe}. At low temperature, only the particle-like band is populated and our results agree with previous work. However, at slightly elevated temperatures, a significant population of the light-like band leads to modified angular emission patterns and lifetimes. Clear experimental fingerprints are predicted and explained by a simple four-state model incorporating spin-orbit as well as intervalley exchange coupling.
\end{abstract}

%Use showkeys class option if keyword
\maketitle

%\tableofcontents
\section{Introduction}
Two-dimensional (2D) transition metal dichalcogenides are promising materials for ultra-thin electronic, optoelectronic, photocatalytic and photovoltaic devices \cite{article:optoIntro1, article:optoIntro2,article:FETWS2,article:photoCatalytic,article:photoVoltaic, article:dissociationMassicotte}. In particular, semiconducting monolayer {\MoS}, {\MoSe}, {\WS}, and {\WSe} are of significant interest due to their direct band gap and strong Coulomb effects \cite{article:TMDS, article:MoS2thickness, article:starkShift, article:dissociationMassicotte} with exciton binding energies as large as $\sim$0.5 eV \cite{article:TMDS, article:MoS2thickness,article:darkExcitonsBSE}. This has prompted intense research into understanding the optical and radiative properties of these materials as reported in both experimental and theoretical work \cite{article:MoS2thickness,article:reviewBoi,article:lifetime_model,article:lifetime_model2, article:starkShift, article:dissociationMassicotte}. Moreover, states in the vicinity of the band gap are dominated by the $d$-orbitals of the transition metal atoms \cite{article:basicInfoTMD}. Hence, the highest valence bands are split by more than 100 meV  due to strong spin-orbit (SO) coupling \cite{article:C2DB}. The hexagonal crystal structure of TMDs means that excitons reside in two inequivalent  $K$ and $K'$ valleys. As a consequence, excitons with a finite center of mass momentum follow a highly unusual energy dispersion. That is, $K$ and $K'$ states couple to produce light-like and particle-like bands that are approximately linear and parabolic, respectively \cite{article:momentumArticle1, article:momentumArticle2}. While this unusual dispersion has been firmly established in previous theoretical \cite{article:momentumArticle1, article:momentumArticle2} and experimental \cite{article:EELS-Mom, article:twistedLight, article:Schneider_2020} work, we demonstrate in the present paper that equally striking consequences are expected for light emission from TMD excitons. To this end, we apply \textit{ab initio} modeling to describe effects of center of mass motion on exciton energies, optical matrix elements and light emission. In particular, the thermal population of bands is shown to affect lifetimes and angular emission patterns. Moreover, we demonstrate that an effective four-state model incorporating SO as well as inter- and intravalley exchange coupling successfully captures the essential physics. 

\begin{figure}[ht]
    \centering
    \includegraphics[width=\linewidth]{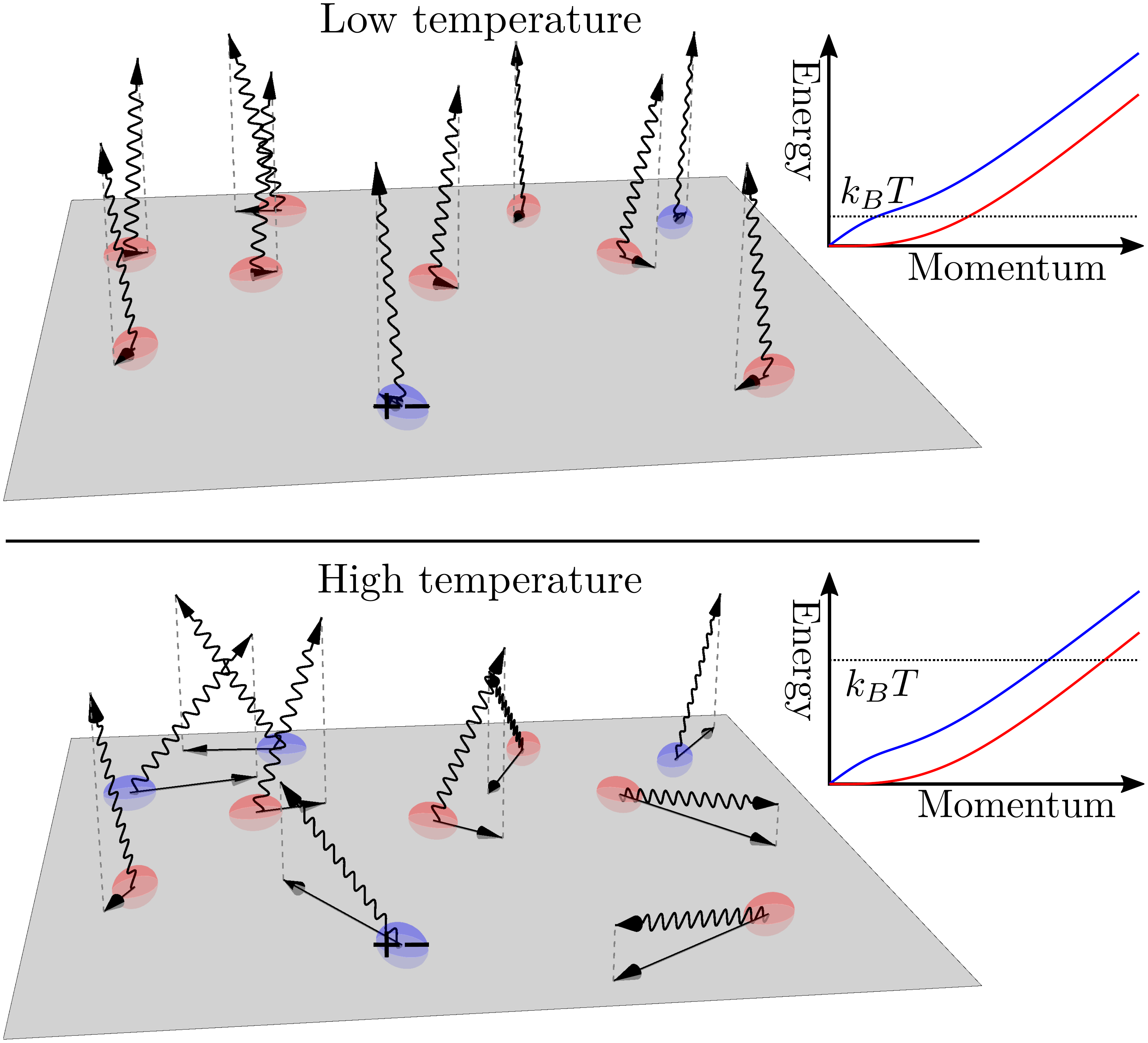}
    \caption{Schematic of photon emission by ensembles of light-like (blue) and particle-like (red) excitons. Photon and exciton momenta are shown as wiggly and straight arrows, respectively. At low temperature, particle-like excitons dominate, whereas nearly identical populations are expected at high temperature. Insets to the right show their energy dispersions.}
    \label{fig:intro2}
\end{figure}

The distinct radiative properties of light- and particle-like TMD excitons are readily understood from simple physical principles. Primarily, whenever an exciton decays by emission of a photon, momentum conservation dictates that the in-plane exciton momentum $\vec{Q}$ is transferred to the photon. In particular, slow and fast excitons emit photons in normal and oblique directions, respectively, as illustrated in Fig. \ref{fig:intro2}. It follows that the thermal population of states with different $\vec{Q}$ is responsible for the emission pattern. The $K-K'$ symmetry of TMDs requires light- and particle-like excitons to be degenerate at $\vec{Q}=0$. However, at finite $\vec{Q}$ two distinct bands are formed as seen in the insets in Fig. \ref{fig:intro2}, with approximately parabolic and linear dispersion at small $\vec{Q}$. Accordingly, the kinetic energy of particle-like states with parabolic dispersions is much lower than that of light-like species at small $\vec{Q}$. In turn, a thermal population favouring particle-like states is expected at low temperature $T$ while near equal populations should exist at elevated temperatures. The energy difference between the bands is due to exchange effects on the order of 10 meV. It follows that an estimate of the characteristic temperature is around 100 K. Accordingly, significantly different exciton populations are expected at low and room temperature.  

A secondary effect is related to the different rates of spontaneous emission of the two exciton species. A high emission rate compared to competing non-radiative mechanisms is crucial for efficient emission. Fundamentally, the rate of spontaneous emission is determined by two factors: transition dipole moment and available phase space. The transition dipole moment is an intrinsically quantum mechanical quantity proportional to the exciton transition momentum matrix element $\vec{P}(\vec{Q})$. We note that the term ''momentum'' appears in two distinct contexts: the center of mass momentum $\vec{Q}$ that is simply a conserved quantum number and the momentum matrix element $\vec{P}$ that is a transition amplitude $\braket{0|\vop{P}|n}$ taken between the ground state $0$ and $n$'th exciton state. To avoid confusion, we will henceforth denote  $\vec{P}$ by the transition matrix element. An accurate calculation of $\vec{P}$ requires inclusion of exciton Coulomb and exchange effects. In the present work, an \textit{ab initio} Bethe-Salpeter equation (BSE) approach combined with accurate density-functional theory (DFT) band structures are applied to this end. The available phase space refers to the result that photon momenta $\vec{q}$ are tightly restricted by geometrical effects. As mentioned above, the in-plane photon momentum is directly determined by the exciton momentum. In turn, under emission into free space, the out-of-plane component is uniquely specified by the photon dispersion relation $q=\omega/c$. Here, the photon frequency $\omega$ follows from energy conservation, i.e. $\hbar\omega=E(\vec{Q})$, where $E(\vec{Q})$ is the exciton energy. Together, these relations impose severe restrictions on the photon modes available for emission. For instance, no emission can take place if $Q>\omega/c$. In fact, phase space is restricted even further by the requirement of photon transversality. This requirement means that emission rates are proportional to $\sin^2\varphi$, where $\varphi$ is the angle between $\vec{q}$ and $\vec{P}$. We demonstrate in the present work that light-like and particle-like excitons are characterised by transition matrix element that are almost entirely parallel and perpendicular to $\vec{Q}$, respectively. Hence, different available phase spaces are found for the two exciton bands leading to differences in their emission rates. Strikingly, at grazing angles $Q\sim\omega/c$, emission from light-like excitons is forbidden while particle-like states remain fully bright. We evaluate and analyse all these effects below and provide concrete suggestions and predictions for experimental verification of our findings via the angular and thermal behaviour of emission spectra.  

%\begin{figure*}[ht]
%    \centering
%    \includegraphics[trim={6cm 3.5cm 9cm 6.5cm},clip,width=\linewidth]{Illustrations/Energies.eps}
%    \caption{Dispersion plots for the lowest energy bright excitons for {\MoS}, {\MoSe}, {\WS}, {\WSe}, as well as for differently screened {\MoS}, where the dielectric constants $\varepsilon$ chosen describes either the screening of a SiO$_2$ substrate ($\varepsilon = 1.55$) or the screening from encapsulation in hBN ($\varepsilon = 4.5$). Furthermore, the magnitude of the transition matrix element projected upon $\vec{Q}$ has been plotted as a vector from each point, with components corresponding to either parallel or perpendicular projection.}
%    \label{fig:dispersions}
%\end{figure*}

Exciton band structures in TMDs have previously been described within the DFT+BSE framework \cite{article:momentumArticle1, article:momentumArticle2} and a similar approach is applied here. In particular, we use a strictly two-dimensional (2D) model of exciton screening incorporated via the Keldysh potential \cite{article:keldysh}. Such a 2D approach has been found to successfully describe particle- and light-like bands \cite{article:momentumArticle2}. Note, though, that the approach by Deilmann \textit{et al.} \cite{article:momentumArticle1} adopts a 3D screening model, and consequently does not observe the light-like band. While the 3D screening approach is justifiable for TMDs in a 3D environment \cite{article:Pedersen_2016}, it fails to describe intrinsic 2D properties because of the applied truncation of Coulomb interactions in the perpendicular direction to eliminate unphysical coupling between periodically repeated layers \cite{article:momentumArticle1}. We calculate  $\vec{Q}$-dependent transition matrix elements that, in turn, yield the decay rate of emission via Fermi's golden rule.  Subsequently, we thermally average over the exciton bands to obtain average lifetimes, yielding the temperature dependence of both lifetime and emission profiles. A similar approach was applied by Palummo \textit{et al.} for calculating the radiative lifetimes of the four TMDs \cite{article:lifetime_model}. However, their model is based on an effective mass approximation for the exciton dispersion, meaning that only the particle-like band is included. These authors find room temperature radiative lifetimes in the few hundreds of ps range, significantly lower than experiments. Longer radiative lifetimes have been obtained by Wang \textit{et al.} \cite{article:lifetime_model2}, modelling the radiative lifetimes of excitons and trions using a modified Wannier-Mott approach neglecting, however, long range exchange interaction between the $K$- and $K'$-valleys \cite{article:lifetime_model2} and, hence, only including the particle-like band. Still, these authors predict radiative exciton lifetimes for {\MoS} to be about one nanosecond at room temperature. In addition, the radiative exciton lifetimes follow a quasi-linear temperature dependence in agreement with the present results, provided we only consider particle-like states. In the present work, we apply a highly accurate DFT+BSE approach to determine the role of light-like bands in radiative exciton lifetimes. We show that a nonlinear temperature dependence of the emission is expected due to the different thermal populations of particle- and light-like bands. Moreover, the unique band structure produces clear fingerprints in the angular spectrum of radiated light. These, for instance, become evident when comparing spectra at low and high temperatures. Finally, our effective four-state model clarifies the important roles of SO splitting and the inter- and intravalley exchange couplings. In particular, emission is found to be highly sensitive to the competition between SO and exchange couplings.

\begin{figure*}[ht]
    \centering
    \includegraphics[width=\linewidth]{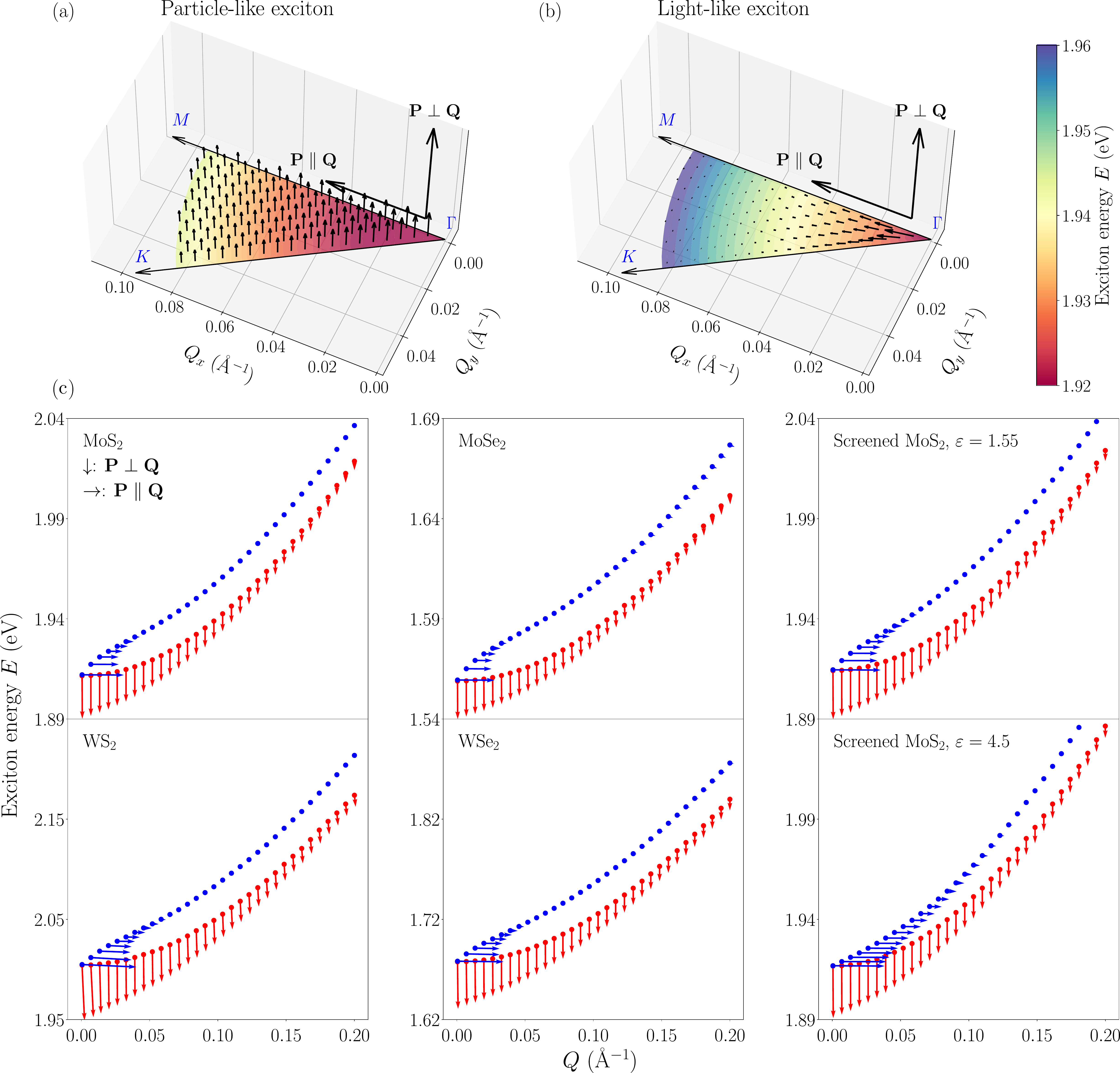}
    %\caption{(a) and (b) exciton band structure for the two lowest bright excitons of {\MoS} vs. $\vec{Q}$, as well as the absolute magnitude of the exciton transition matrix elements projected upon $\vec{Q}$, shown by the size and direction of the arrows. (a) and (b) shows a linear and parabolic $Q$-dependence for the energy, as well as a perpendicular and parallel projection for the exciton transition matrix element respectively. (c) exciton band structures along the $\Gamma \rightarrow M$ high symmetry line for the two lowest bright excitons of {\MoS}, {\MoSe}, {\WS}, {\WSe}, as well as for differently screened {\MoS}. The dielectric constants $\varepsilon$ chosen describes either the screening of a SiO$_2$ substrate ($\varepsilon = 1.55$) or the screening from encapsulation in hBN ($\varepsilon = 4.5$). Furthermore, the magnitude of the transition matrix element projected upon $\vec{Q}$ has been plotted as a vector from each point, with components corresponding to either parallel or perpendicular projection.}
    \caption{Exciton band structure for the two lowest bright excitons vs. $\Vec{Q}$ with exciton transition matrix elements  $\vec{P}$ shown by arrows. Horizontal and vertical arrows indicate matrix elements parallel and perpendicular to $\Vec{Q}$, respectively. Bands as a function of $\Vec{Q}$ in the entire irreducible Brillouin zone for {\MoS} (a) and (b). Particle-like (red) and light-like (blue) bands in {\MoS}, {\MoSe}, {\WS}, {\WSe}, and two examples of screened {\MoS} as a function of $\Vec{Q}$ along the $\Gamma \rightarrow M$ high symmetry line (c). The dielectric constant $\varepsilon$ describes either the screening from a SiO$_2$ substrate ($\varepsilon = 1.55$) or from encapsulation in hBN ($\varepsilon = 4.5$).}
    \label{fig:2D_dispersion}
\end{figure*}

\section{Exciton Band Structures}
\label{sec:momentum}
First-principles BSE exciton band structures are numerically demanding because resolution corresponding to optical momenta is required. For this reason, we have developed a computationally efficient approach, in which a limited set of bands is retained and interpolation between $\Vec{Q}$-points is applied to acquire exciton energies and transition matrix elements. Briefly, we solve the BSE based on DFT band structures obtained using the GPAW-package \cite{article:GPAW1, article:GPAW2, article:GPAW3, article:GPAW4, article:GPAW5}. We scissor-shift the DFT band gap, such that the lowest bright excitation matches the $A$-peak of the experimental spectrum measured by Hsu \textit{et al.} \cite{article:MoSe2refractive}. The BSE is then constructed from four valence- and four conduction bands. Band states are two-component spinors due to SO coupling but we define the BSE matrix in a basis of definite spin projections $n_{\vec{k},\sigma}$ with band index $n$, wave vector $\vec{k}$ and spin $\sigma$. The BSE matrix is constructed for a $60 \times 60$ $k$-point grid covering the full Brillouin zone. We apply the Keldysh potential \cite{article:keldysh} for the Coulomb interaction, such that the potential in Fourier  $\Vec{q}$-space is modelled as $w_C(\Vec{q}) = v_{\text{bare}}(\Vec{q})/(\eps + r_0 q)$ where $v_{\text{bare}}(\Vec{q})$ is the bare 2D Coulomb interaction in Fourier space, $\varepsilon$ the external dielectric screening found by averaging sub- and superstrate dielectric constants, and $r_0$ the screening length given as $r_0 = 2 \pi \chi_{0,xy}$. Here, $\chi_{0,xy}$ is the in-plane static sheet polarisability obtained directly from the DFT calculation. This approach has proven effective at describing the screening of the Coulomb interaction in TMDs \cite{article:screeningTrolle,article:MoS2thickness}. Furthermore, only external charges screen the exchange term that is given as $w_x(\Vec{q}) = v_{\text{bare}}(\Vec{q})/\eps$ \cite{article:exchangeScreening,article:Louie_2021_exchange}. Finally, the center of mass momentum $\Vec{Q}$ is handled by adding it to the $\Vec{k}$-vector such that a conduction band state $c_{\vec{k},\sigma}$ is replaced by $c_{\vec{k}+\Vec{Q},\sigma}$, while the valence band states $v_{\vec{k},\sigma}$ are unchanged. However, momenta $\vec{k}$ and $\Vec{Q}$ will be restricted to the same grid unless additional DFT calculations for wave vectors $\Vec{k}+\Vec{Q}$ are made for every $\Vec{Q}$ not in the $k$-point grid. Thus, to increase resolution to the optical momentum range, a $60 \times 60$ conduction band $k$-grid shifted by $\Vec{Q}$ is obtained by interpolation from a fixed $120 \times 120$ point dense grid, using a 4th-degree polynomial spline scheme. This approach ensures that any momentum $\Vec{Q}$ can be resolved, without the need for additional expensive DFT calculations. As a consequence of the added center of mass momentum, the transition matrix elements are calculated as 
\begin{equation}
    \vec{P} = \frac{A}{4 \pi^2}\sum_{v,c,\sigma} \int \Psi_{v,\vec{k},\sigma}^{c,\vec{k}+\Vec{Q},\sigma} \braket{v_{\Vec{k},\sigma}|\vop{p}(\vec{q})|c_{\Vec{k} + \Vec{Q},\sigma}}d^2 \vec{k},
    \label{eq:momElems}
\end{equation}
where $A$ is the TMD area, $\Psi_{v,\vec{k},\sigma}^{c,\vec{k}+\Vec{Q},\sigma'}$ the exciton wave function in the basis of valence-conduction band transitions with definite spin projection $\sigma,\sigma' \in \{ \uparrow, \downarrow \}$, and $\Vec{q}$ the photon momentum. Also, $\vop{p}(\vec{q})$ is the momentum operator given as
\begin{equation}
    \vop{p}(\vec{q}) = -\frac{i \hbar }{2} \left\{ \nabla \exp{-i \Vec{q} \cdot \Vec{r}}  + \exp{-i \Vec{q} \cdot \Vec{r}} \nabla \right\},
\end{equation}
and we exploited the fact that matrix elements of $\vop{p}$ are diagonal in spin.

% Additional Details
%The pseudo wave function are given in a plane-wave expansion, and the DFT calculations have been carefully checked for convergence of the band structures, wave functions, and binding energy of the subsequent BSE calculation.  In particular a plane-wave cutoff-energy of 600 eV is chosen, and the BSE calculations are performed on a $60 \times 60$ $k$-point Monkhorst grid. have been developed in-house and handles the exciton band structures by adding a momentum $\Vec{Q}$ to every conduction state $\Ket{c, \vec{k}} \rightarrow \ket{c, \vec{k}+\vec{Q}}$, yielding a set of energies and wave functions for every momentum $\vec{Q}$. These are then in turn used to calculate the transition matrix elements. The addition of a center of mass momentum to the conduction state, will in effect require a separate DFT calculation for every $\vec{Q}$ not included in the $k$-point grid. However, to calculate the exciton dispersion in a reasonable time-frame, an interpolation scheme has been developed for the DFT, circumventing this requirement. To ensure reliable interpolation results, the DFT interpolant grid was chosen to be a $120 \times 120$ $k$-point Monkhorst grid.

Generally, exciton transition matrix elements depend on both the magnitude and direction of $\Vec{Q}$. In Fig. \ref{fig:2D_dispersion}a and b, this dependence is shown for the particle- and light-like bands of {\MoS}, respectively. Fortunately, as demonstrated by Fig. \ref{fig:2D_dispersion}a and b, there is little angular dependence for both energy dispersion and transition matrix elements within the thermally allowed $Q$-range at room temperature. This is a consequence of the threefold symmetry of TMDs, which for low values of $Q$ deviates only slightly from circular symmetry, allowing us to significantly simplify the calculations of the emission properties by only requiring $\vec{Q}$ in one dimension. Furthermore, it can be observed that the transition matrix elements are always practically perpendicular to the $\vec{Q}$-vector in the lower energy band, and parallel to the $\vec{Q}$-vector in the upper energy band. This is a consequence of hexagonal symmetry in TMDs. In Sec. \ref{sec:model}, we present a rigorous analysis of the symmetry properties including exchange and SO coupling. However, a very simple physical picture can be constructed if SO coupling is temporarily ignored. Firstly, at $\Vec{Q}=0$, the $K/K'$-valley excitons $\ket{K}$ and $\ket{K'}$ are degenerate as required by time reversal symmetry. At finite $\vec{Q}$, the degeneracy is lifted by intervalley exchange $v_x$. To first-order in $\vec{k}\cdot\vec{p}$ perturbation theory, intervalley exchange is (see App. A)
\begin{equation}
    \braket{K|v_x|K'} \propto w_x(Q)  (\vec{Q} \cdot \vec{p}_{\vec{K}}^*) (\vec{Q} \cdot \vec{p}_{\vec{K'}}),
    \label{eq:exc_propto}
\end{equation}
where $\vec{p}_{\vec{k}}=\braket{v_{\Vec{k},\sigma}|\vop{p}(0)|c_{\Vec{k},\sigma}}$ is an interband transition matrix element. Because $w_x(Q) \propto 1/Q$, light-like energies increasing by $2|\braket{K|v_x|K'}|$ form a linear exciton band while particle-like energies are unperturbed to first order. This essentially explains the characteristic band structure. Next, the approximate circular symmetry around the $K/K'$-points allows us to choose $\vec{Q}$ along $x$. If we introduce the chiral unit vectors $\unvec{e}_{\pm}\equiv(\unvec{x}\pm i\unvec{y})/\sqrt{2}$ then, with an appropriate choice of phase, $\vec{p}_{\Vec{K}}=p_0 \unvec{e}_-$ and $\vec{p}_{\vec{K'}}=-p_0 \unvec{e}_+$, where the minus sign is dictated by time reversal symmetry and $p_0$ is real-valued \cite{article:selectionRules}. Note that choosing $\vec{Q}=Q\unvec{x}$ makes intervalley exchange coupling real-valued as well, so that the coupled states are simply $\ket{p}=(\ket{K}+\ket{K'})/\sqrt{2}$ and $\ket{l}=(\ket{K}-\ket{K'})/\sqrt{2}$ for particle- and light-like states, respectively. It follows immediately that $\vec{P}_p\propto\vec{p}_{\Vec{K}}+\vec{p}_{\Vec{K'}}$ and $\vec{P}_l\propto\vec{p}_{\Vec{K}}-\vec{p}_{\Vec{K'}}$, which are clearly parallel and perpendicular to $\vec{Q}$, accordingly. Thus, the distinct transition matrix elements for the two bands are explained by simple symmetry arguments. In Sec. \ref{sec:model}, we demonstrate that identical conclusions are reached even if SO coupling is retained.  This important conclusion has also been verified by direct comparison to the numerical BSE solution for {\MoS}. In addition, due to the identical parity of the valence band maximum (VBM) and conduction band minimum (CBM), the $z$-components of the transition matrix elements are approximately zero, only differing slightly due to the out-of-plane photon momentum. 

In Fig. \ref{fig:2D_dispersion}c, we show exciton bands along $\Gamma \rightarrow M$ for the two lowest bright states in the four TMDs studied, including two cases of {\MoS} screened by a dielectric environment. The particle- and light-like bands are readily identified, and once again show either a parallel or perpendicular projection of the transition matrix elements on the $\vec{Q}$-vector. In addition, the difference in energy between the two bands increases linearly for small values of $Q$, and quickly settles at 10-20 meV around $Q \simeq 0.03$ $\text{Å}^{-1}$. The magnitude of the transition matrix elements for the light-like band decreases faster with increasing $Q$ than the particle-like band. In particular, this happens when the energy of the light-like exciton approaches the energies of higher energy exciton states, due to the intervalley exchange. For the particle-like parabolic band, the exciton effective mass can be extracted from from a polynomial fit as seen in Table \ref{tab:my_label}. For free-standing {\MoS}, our results are in good agreement with previous values \cite{article:momentumArticle1,article:momentumArticle2}. Figure \ref{fig:2D_dispersion}c also shows two cases of screened {\MoS}, namely with dielectric constants corresponding to quartz substrates and hBN encapsulation. Notably, it can be seen that the transition matrix elements increase slightly in a screened environment. This is counter-intuitive because a decreased electron-hole overlap is expected to reduce oscillator strength within a Wannier model. However, as demonstrated in the effective model presented in Sec. \ref{sec:model}, the effect is caused by the screening of the exchange interaction, which is not accounted for in the Wannier picture. Moreover, an additional consequence of the exchange screening is a significant reduction of the slope for the light-like band.

\begin{figure}[ht]
    \centering
    \includegraphics[trim={0.9cm 0.0cm 1.5cm 0cm},clip,width=\linewidth]{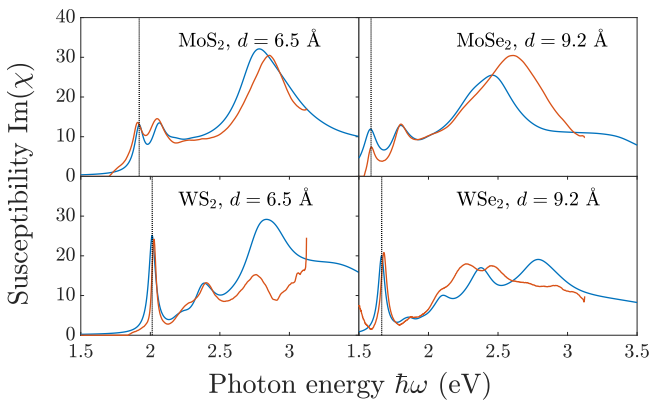}
    \caption{Imaginary part of the susceptibility calculated from the transition matrix elements (blue) compared to experiment by Hsu \textit{et al.} (red) \cite{article:MoSe2refractive} for TMDs on SiO$_2$ ($\varepsilon = 1.55$). An energy dependent broadening of $\hbar \Gamma \propto \hbar \omega - E_A$ is applied, with the energy of the first bright exciton $E_A$ indicated by dotted lines. The parameter $d$ is the material thickness used to transform between computed 2D and measured 3D results.}
    \label{fig:experimental}
\end{figure}

In the following section, exciton transition matrix elements are applied to compute radiative decay rates emphasizing the role of the exciton band structure. As an additional  test of our computational approach, we compare in Fig. \ref{fig:experimental} the electric susceptibilities of the TMDs, as calculated from the BSE solution at $\vec{Q} = 0$ using the momentum approach \cite{article:ExcitonGauge} and as measured by Hsu \textit{et al.} \cite{article:MoSe2refractive}. In these experiments, TMDs are placed on quartz (SiO$_2$) substrates, which has proven to strongly affect both exciton binding energies and quasiparticle energies \cite{article:screeningTrolle}. Thus, the BSE is solved with an applied screening of $\varepsilon = \frac{2.1+1}{2} = 1.55$, where $2.1$ is the SiO$_2$ dielectric constant. A good agreement, both in energy and amplitude, between calculation and experiment is observed, in particular, around the two low-energy $A$- and $B$-peaks. Note, though, that by construction the calculated spectra must agree with the $A$-peak since this requirement is used to fix the quasiparticle scissors-shift. The discrepancy at higher energy originates mainly from the omission of high-energy bands. The figure also illustrates the significant SO splitting of the four TMDs, which means that the exciton radiative lifetime will be dominated by the $A$-exciton. This is particularly true for tungsten-based TMDs, where the SO coupling is so significant that the $B$-exciton lies within the $A$-exciton continuum.

\section{Radiative Lifetimes}

\begin{table*}[htbp]
    \centering
    \caption{Calculated properties of TMDs and comparison to experimental results. Here, $\tau_{300\text{K}}$ and $\tau_{30\text{K}}$ are the calculated radiative lifetimes at room temperature and 30K in vacuum, $\tau_{300\text{K},\text{exp}}$ the experimental room temperature lifetime, and $M_p$ the effective mass of the particle-like exciton band. Experimental conditions: *SiO$_2$ substrate, \dag Superacid defect passivation with oleic acid (OA) or bis(trifluoromethane)sulfonimide (TFSI), \ddag Extrapolated from $77$K to room temperature.}
    \label{tab:my_label}
    \begin{tabular}{M{2cm}|M{3cm}M{3cm}M{4.5cm}M{3cm}}
        & {\MoS} & {\MoSe} & {\WS} & {\WSe} \\
        \hline
        $\tau_{300\text{K}}$ (ns) & 1.18 & 1.95 & 0.35 & 0.62 \\
        $\tau_{300\text{K},\text{exp}}$ (ns) & 0.58 (TFSI) *\dag\ddag \cite{article:darkExc} & 1* \cite{article:intrinsicRadiative} & 0.25 (OA) or 0.31 (TFSI) *\dag \cite{article:passivatedWS2} & 0.6* \cite{article:WSe2-Data}\\
        $\tau_{30\text{K}}$ (ps) & 80.7 & 139 & 25.6 & 47.0 \\
        $M_p (m_e)$ & 1.40 & 1.53 & 0.867 & 0.877
    \end{tabular}
\end{table*}

\begin{figure*}[ht]
    \centering
    \subfloat[\label{fig:emission:lifetime}]{
    \includegraphics[trim={0.5cm 0.0cm 2.0cm 3.5cm},width=\columnwidth]{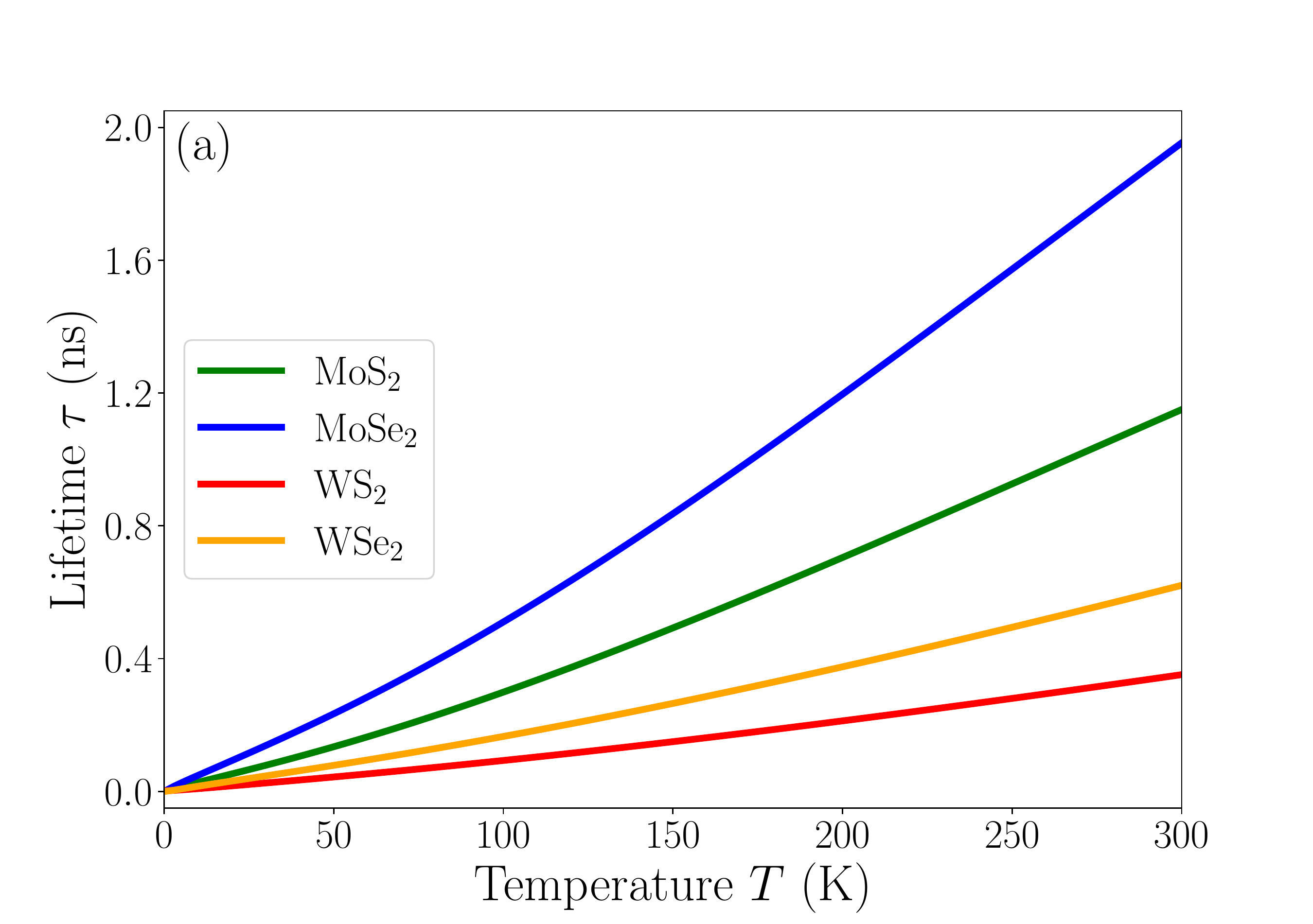}
    }
    \subfloat[\label{fig:emission:diff}]{
    \includegraphics[trim={0.5cm 0.0cm 2.0cm 3.5cm},width=\columnwidth]{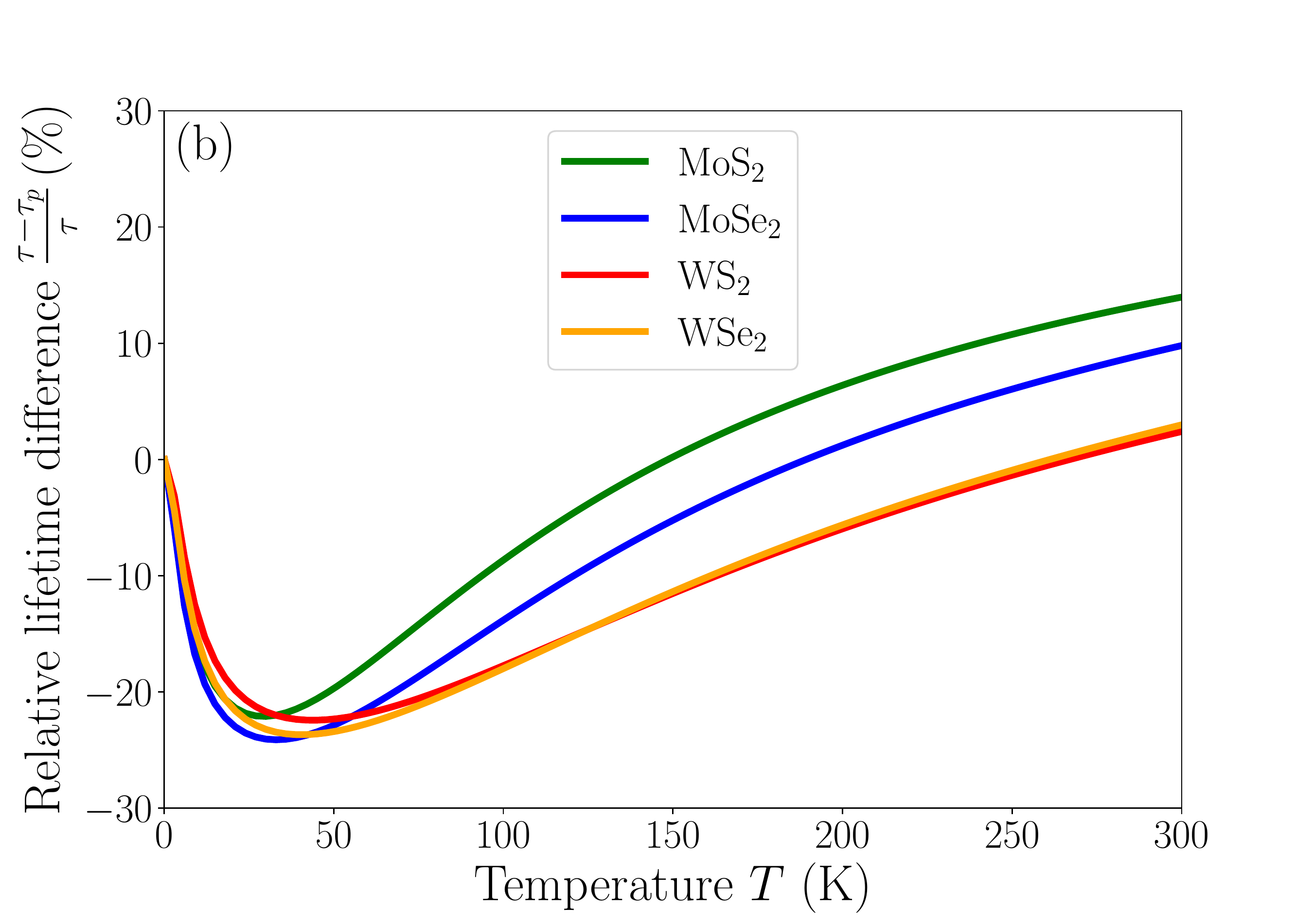}
    }\\
    \subfloat[\label{fig:emission:spectrum}]{
    \includegraphics[trim={0.5cm 0.0cm 2.0cm 2.7cm},width=\columnwidth]{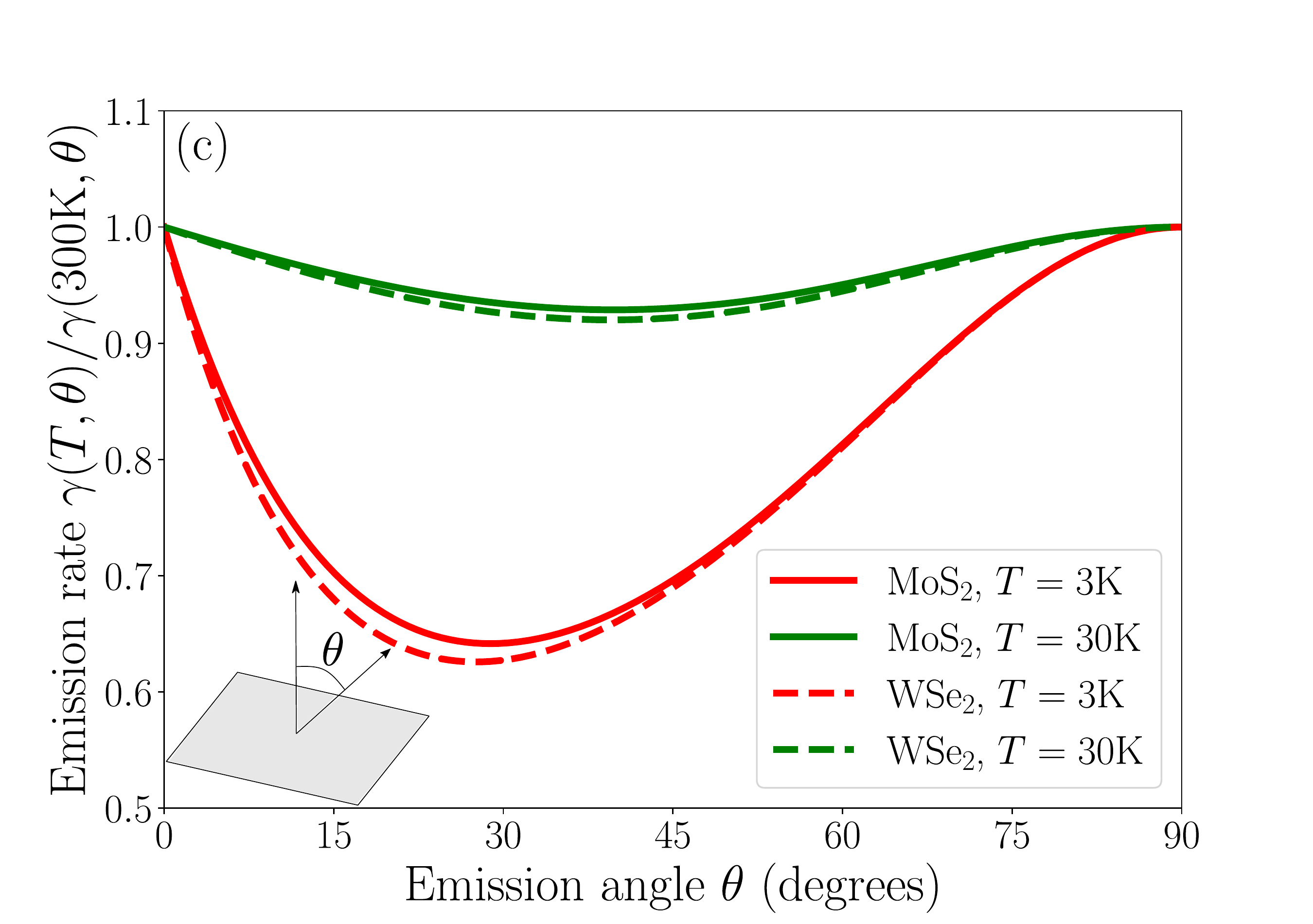}
    }
    \subfloat[\label{fig:emission:angle}]{
    \includegraphics[trim={0.5cm 0.0cm 2.0cm 2.7cm},width=\columnwidth]{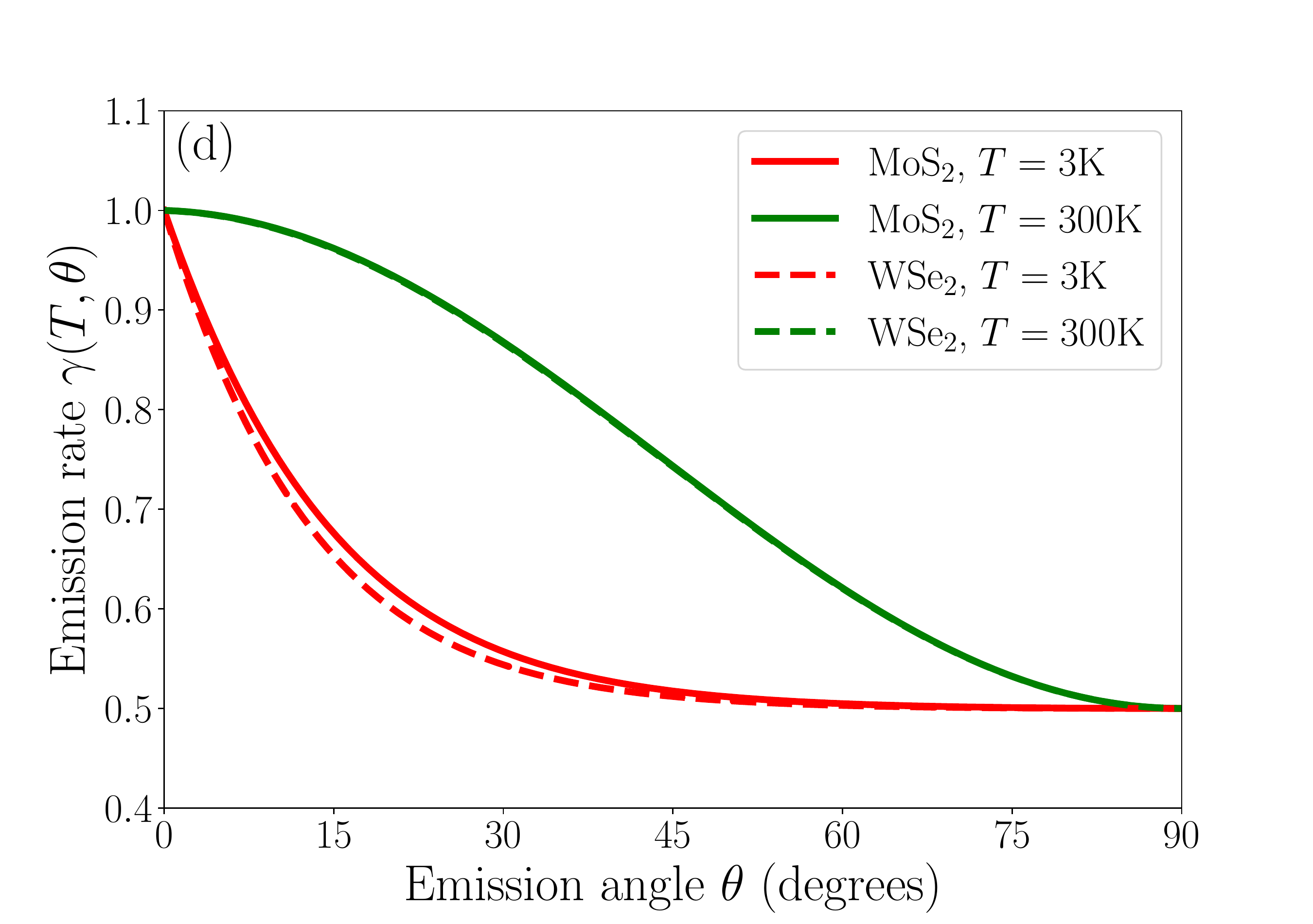}
    }
    \caption{(a) Calculated radiative lifetimes for four TMDs as a function of temperature. (b) Relative difference between lifetimes obtained from the particle-like band alone $\tau_p$ and the full calculation $\tau$. (c) Normalised emission $\gamma(T,\theta)/\gamma(300\text{K},\theta)$ at temperatures $T = 3$K and $T = 30$K as a function of the emission angle $\theta$. (d) Emission rate $\gamma(T,\theta)$ as a function of $\theta$ for low and high temperature. Here, the characteristic energy of light-like excitons $\hbar V\omega/c$ is 0.973 meV and 1.111 meV for {\MoS} and {\WSe}, respectively.}
    \label{fig:emission}
\end{figure*}

%To develop a model for the radiative lifetimes, based on the momentum resolved exciton properties, we start at Fermi's Golden rule. %One formulation of this is as follows:
%\begin{equation}
%    \Gamma=\frac{2\pi}{\hbar}\sum_f\norm{\braket{f|\op{\mathcal{H}}_{int}|i}}^2\delta(E_{fi}),
%\end{equation}
Radiative lifetimes are modelled using Fermi's golden rule \cite{article:lifetimeEquation}, which yields a momentum-resolved radiative rate for 2D materials given by
\begin{equation}
\begin{split}
    \Gamma(\vec{Q})&=\frac{e^2}{2m_e^2A\eps_0\hbar\omega c}\sum_{q_x,q_y}\delta_{q_x,Q_x}\delta_{q_y,Q_y} \\ &\times \int{\vec{P}^*\cdot(\vec{I}-\unvec{q}\unvec{q})\cdot\vec{P}}\,\delta\left(q-\textstyle\frac{\omega}{c}\right)\,dq_z.
\end{split}
\label{eq:Gamma}
\end{equation}
Here, the Kronecker deltas ensure in-plane momentum conservation and the Dirac delta function $\delta\left(q-\textstyle\frac{\omega}{c}\right)$ arises from the requirement of energy conservation. Defining the emission angle $\theta$ as the angle to the surface normal, the exciton and photon momenta in the light cone are related via $Q=q\sin\theta$ and $q_z=q\cos\theta$. Furthermore, Eq. \eqref{eq:Gamma} can be generalised to a 2D emitter encapsulated in a linear isotropic material, by replacing $\eps_0 \rightarrow n \eps_0$ and $\omega / c \rightarrow \omega n / c$ in the Dirac delta function, with $n$ being the refractive index of the medium. The exciton transition matrix elements $\vec{P}$ are calculated from the BSE eigenstates using Eq. \eqref{eq:momElems}. The integral is evaluated for exciton transition matrix elements either parallel $\vec{P} \parallel \Vec{Q}$ or perpendicular $\vec{P} \perp \Vec{Q}$, thus yielding two cases
\begin{equation}
\begin{split}
    \Gamma(\Vec{Q}) &= \Gamma_0(\Vec{Q}) \begin{cases} 
        \int_Q^{\infty} \frac{q_z}{q} \delta \left(q - \frac{\omega}{c}\right) d q & \vec{P} \parallel \Vec{Q} \\
        \int_Q^{\infty} \frac{q}{q_z} \delta \left(q - \frac{\omega}{c}\right) d q & \vec{P} \perp \Vec{Q}
    \end{cases} \\
    &=  \Gamma_0(\Vec{Q}) \Theta\left( \textstyle\frac{\omega}{c} - Q \right) \begin{cases} 
        \cos\theta & \vec{P} \parallel \Vec{Q} \\
        1/\cos\theta & \vec{P} \perp \Vec{Q}
    \end{cases},
\end{split}
\end{equation}
where $\Gamma_0(\Vec{Q}) = \frac{e^2 ||\vec{P}(\Vec{Q})||^2}{m_e^2A\eps_0\hbar\omega c}$ and $\Theta(x)$ is the Heaviside step function. The first and second of these cases apply to the light- and particle-like bands, respectively. Finally, by assuming a quasi-thermal distribution of excitons, and thus thermally averaging this result, one arrives at an expression for the intrinsic mean radiative rate
\begin{equation}
\label{eq:gammaMean}
    \left< \Gamma \right> = \frac{1}{\sum_n Z_n} \sum_n \sum_{|\vec{Q}| < \omega / c} \Gamma_n(\Vec{Q}) \exp{-E_n(\Vec{Q}) / k_B T},
\end{equation}
where $Z_n$ is the partition function and $E_n(\Vec{Q})$ the exciton energy of state $n$ with center of mass momentum $\Vec{Q}$. Here, we limit $n$ to include only the bright particle- and light-like excitons ($n=p,l$) since radiation does not directly interact with the dark excitons, and thus they are not expected to be significantly populated. In the thermal average, the $Q$-dependent quantities are the exciton energy $E_n(\vec{Q})$ and transition matrix element magnitude $P_n(\vec{Q})$, where the latter only varies very slightly within the light cone. We denote the common energy at the bottom of the bands by $E_0$. Hence, summing only over the two bands and approximating $\Gamma_0(\Vec{Q})\approx \Gamma_0(0)$, $E_p(\vec{Q})\approx E_0$ and $E_l(\vec{Q})\approx E_0+\hbar V Q$ with velocity $V$, we find
\begin{equation}
\label{eq:gammaAngular}
    \left< \Gamma \right> \approx \frac{4\pi\omega\Gamma_0(0)}{c(Z_p+Z_l)} \int_0^{\pi/2}\gamma(T,\theta)\sin\theta d\theta,
\end{equation}
where $\gamma(T,\theta)=(1+\cos^2\theta\exp{-\eps_l\sin\theta})/2$ with $\eps_l=\hbar V\omega/(c k_B T)$ and normalised such that $\gamma(T,0)=1$. The angular integral in  Eq. \eqref{eq:gammaAngular} can be evaluated analytically leading to a combination of Bessel  $I_2$ and Struve $\textbf{L}_2$ functions of the form $\pi[\textbf{L}_2(\eps_l)-I_2(\eps_l)]/4\eps_l+2/3$. Note that the energy approximations are not made in the evaluation of the partition function integrals that cover a wide range of $\vec{Q}$-vectors. At low temperature, $\eps_l\gg 1$ and the light-like contribution is minute. Conversely, at high temperature, the contributions from the two bands are comparable. The forbidden emission from light-like states into grazing directions $\theta\approx\pi/2$ is manifest in the $\cos^2\theta$ weight.

% Due to population change between the light-like and particle-like band
The thermally averaged radiative lifetimes $\tau=\left< \Gamma \right>^{-1}$, given as a function of temperature, for the different TMDs can be seen in Fig. \ref{fig:emission:lifetime}. At low temperatures, the lifetime increases roughly linearly with temperature, due to the prevalence of the particle-like band, in agreement with previous theoretical work \cite{article:lifetime_model, article:lifetime_model2}. However, the addition of the light-like band adds a noticeable curvature to the relation, in contrast to earlier work. This is a consequence of the temperature dependent population of the two bands. Furthermore, in Fig. \ref{fig:emission:diff}, we plot the relative difference between the full lifetime, including both particle- and light-like bands, and approximate results omitting the light-like band. The deviations reaching 25 $\%$ provide a quantitative estimate of the error inherent in usual lifetime models considering only particle-like bands \cite{article:lifetime_model, article:lifetime_model2, article:lifetimeEquation}. Finally, light-like states also affect the angular spectrum, as illustrated in Fig. \ref{fig:emission:spectrum} and \ref{fig:emission:angle}. In Fig. \ref{fig:emission:spectrum}, the normalised angular profiles at low and room temperature are clearly distinct, showing deviations of the ratio $\gamma(T,\theta)/\gamma(300\text{K},\theta)$ from one reaching 35 $\%$. The same trend is visible in Fig. \ref{fig:emission:angle}, showing an increased out-of-plane radiative rate at low temperature. Note, however, that low temperature  ($T < 50$K) experiments show the radiative decay of excitons to be much faster than phonon scattering, implying that the quasi-thermal equilibrium assumption is questionable \cite{article:intrinsicRadiative}. 

\begin{figure*}[htbp]
    \centering
    \subfloat[\label{fig:emissionScreened:lifetime}]{
    \includegraphics[trim={0.5cm 0.0cm 2.0cm 3.5cm},width=\columnwidth]{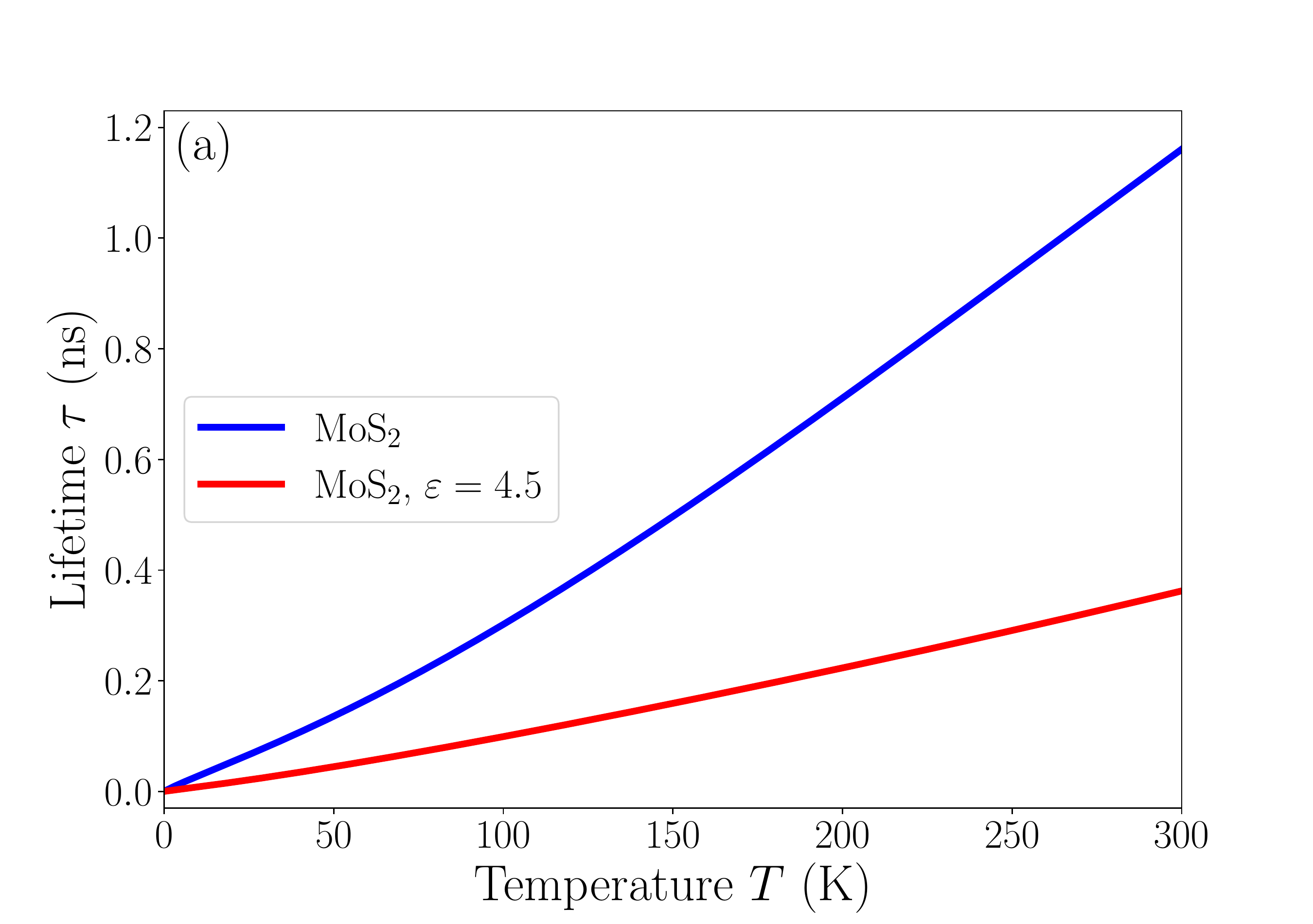}
    }
    \subfloat[\label{fig:emissionScreened:spectrum}]{
    \includegraphics[trim={0.5cm 0.0cm 2.0cm 3.5cm},width=\columnwidth]{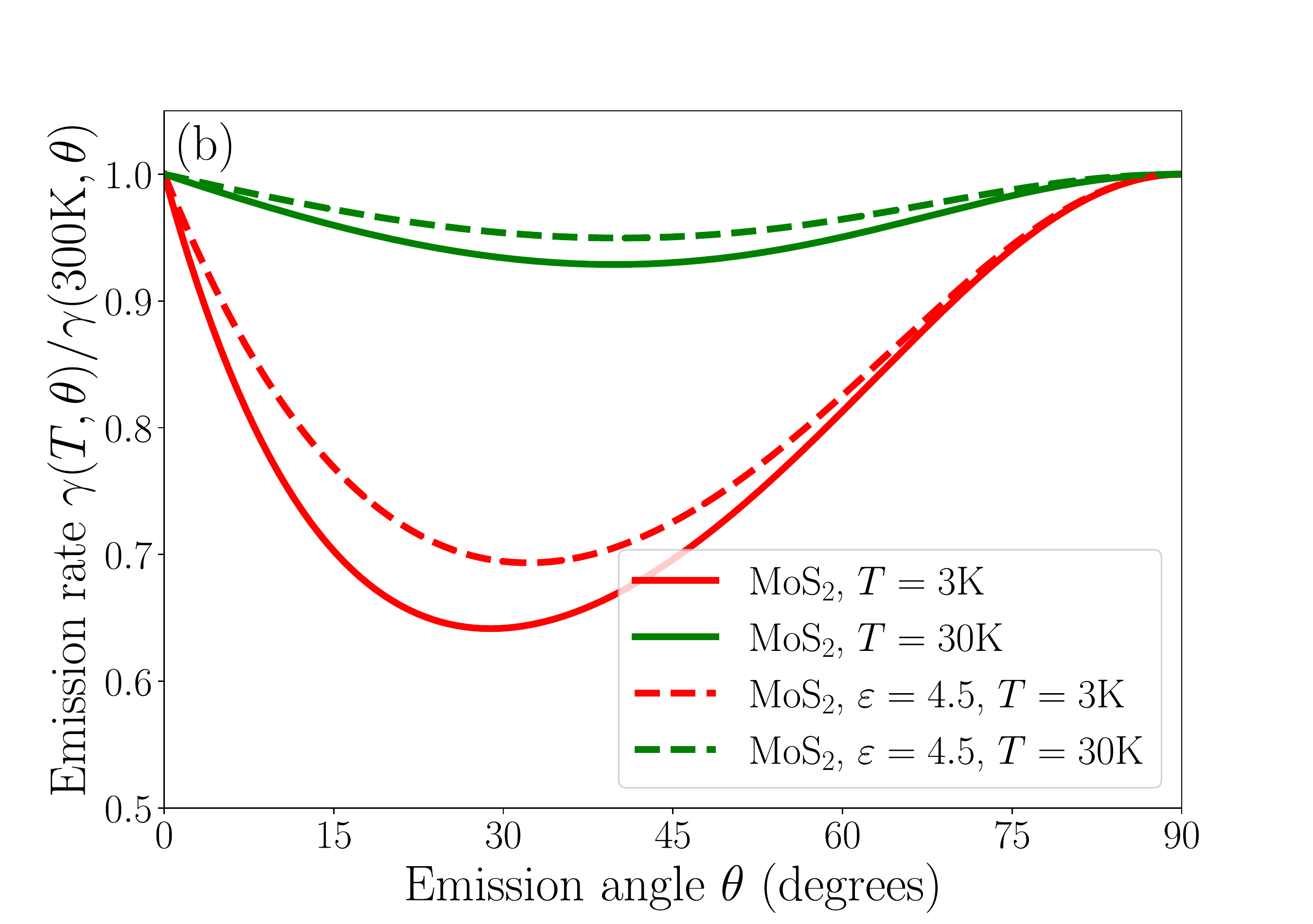}
    }
    \caption{(a) Calculated lifetimes for screened and unscreened {\MoS} as a function of temperature. (b) Normalised emission $\gamma(T,\theta)/\gamma(300\text{K},\theta)$ at $T = 3$K and $T = 30$K as a function of the emission angle $\theta$. The characteristic energy of light-like excitons in hBN encapsulated {\MoS} with $\varepsilon$ = 4.5 is $\hbar V\omega \sqrt{\varepsilon}/c$ = 0.661 meV.}
    \label{fig:emissionScreened}
\end{figure*}

The intrinsic radiative lifetimes for TMDs are notoriously difficult to measure experimentally, making comparison to theoretical predictions difficult. This difficulty stems from numerous effects such as exciton screening by substrates \cite{article:screeningTrolle}, Purcell effects, defect-assisted recombination \cite{article:reviewBoi, article:amani-1, article:amani-2, article:amani-3, article:amani-4, article:amani-5}, and defect-trapped excitons \cite{article:darkExc, article:MaxPassive, article:passivatedWS2}. In particular, defect induced effects mean that significantly varying experimental lifetimes have been reported for the TMDs. For example, the lifetime of {\MoS} emission at $T=300$K has been found ranging from $\tau = 50$ ps to $\tau > 10$ ns \cite{article:reviewBoi}. The usual approach to measuring radiative lifetimes is by fitting to time-resolved photoluminescence measurements. Using this method Shi \textit{et al.} finds the emission lifetime of suspended {\MoS} at room temperature to be $\tau = 850 \pm 48$ ps, and Si$_3$N$_4$ supported {\MoS} to be $\tau = 469 \pm 26$ ps \cite{article:simpleMeasurement}. The ratio between these two lifetimes is approximately equal to the refractive index of Si$_3$N$_4$, and is likely caused by exciton screening and the Purcell effect, providing some insight into the magnitude of these effects. 

Many TMD lifetime measurements are characterised by low quantum yield due to significant defect-assisted Auger recombination caused, in particular, by chalcogenide vacancies. To counteract this, Xu \textit{et al.} fabricated high-quality {\MoS} on sapphire substrates, achieving high photoluminescence yield and a radiative lifetime of $\tau = 930$ ps at room temperature \cite{article:high-quality-mos2}. Other groups have taken a different approach to achieving high quantum yields, by passivation of sulphur vacancies in {\MoS} and {\WS} by either chemical treatment or gating \cite{article:amani-1,article:amani-2,article:amani-3,article:amani-4}. Defect passivation leads to high quantum yields approaching unity, as well as radiative lifetimes in the 10 ns range at room temperature. These very long radiative lifetimes have subsequently been attributed to defect-induced exciton traps in TMDs \cite{article:darkExc, article:MaxPassive}. Goodman \textit{et al.} propose that by saturating these exciton traps, the intrinsic radiative lifetime can be extracted, yielding a lifetime of $\tau = 150$ ps at $77$K for SiO$_2$ supported {\MoS} \cite{article:darkExc}. Assuming a linear temperature dependence, this extrapolates to $\tau = 585$ ps at room temperature. In direct comparison, the present work predicts the lifetime of {\MoS} at $77$K to be $\tau = 221$ ps for suspended {\MoS}. It is striking that the ratio between calculated and measured values is very close to the refractive index of the substrate used in the measurement. A similar trend is observed for our {\MoSe} calculation compared to experiment in Tab. \ref{tab:my_label}, while the tungsten based TMDs seem less affected by the substrate. This discrepancy is presumably a consequence of the larger SO effect in the tungsten based TMDs due to two separate effects: Firstly, the higher ratio between the SO coupling and intravalley exchange yields a brighter exciton, as shown in Sec. \ref{sec:model}, which due to exchange screening affects {\MoS} and {\MoSe} the most. Secondly, in tungsten compounds, the conduction band state midways between the $K$ and $\Gamma$ symmetry points, i.e. the $Q$-valley, is almost degenerate with the one at the $K$-point. BSE calculations by Deilmann \textit{et al.} show that, as a consequence, dark excitons with a hole in the $K$-valley and electron in the $Q$-valley have a lower energy than the $K,K$-valley exciton \cite{article:momentumArticle1}. Since these dark excitons are not included in our quasi-thermal distribution, any actual population of these will increase the radiative lifetimes. In fact, Madéo \textit{et al.} \cite{article:JMadeo} provide direct visualisation of these dark $K,Q$-valley excitons and find a significant population of up to $50\,\%$ of the total excitons. Moreover, the high energy barrier of almost $0.3 \, \text{eV}$ between $K,K$-valley and $K,Q$-valley excitons \cite{article:momentumArticle1} makes populations highly dependent on the excitation method.

%The increased transition matrix elements for the SiO$_2$ substrate can be seen in figure \ref{fig:2D_dispersion}c for {\MoS}. Furthermore, it can be seen that increasing the screening even further to the case of hBN encapsulation, does not significantly increase the transition matrix elements further, due to the significantly reduced Coulomb interaction from the large screening. Furthermore, at low values of $Q$ the exchange interaction between the two bands is diminished, reducing the slope of the light-like band. 
When including the dielectric environment in the lifetime calculations for the hBN encapsulation case, we observe significantly reduced lifetimes, as shown in Fig. \ref{fig:emissionScreened:lifetime}. This is a combination of the exchange screening effect mentioned in Sec. \ref{sec:model}, and the Purcell effect. The latter effectively increases the phase space of optically allowed states according to the substitution $\omega/c\rightarrow n\omega/c$ resulting in significantly more excitonic states coupling to light. This increased phase space also manifests in the angular spectrum of emitted light, as seen in Fig. \ref{fig:emissionScreened:spectrum}, where the angle of minimum emission and thermal dependence have increased and decreased slightly, respectively, compared to the vacuum calculation. The approximations of Eq. \eqref{eq:gammaAngular} have been applied to obtain the angular spectrum. Note, though, that the $E_p \approx E_0$ approximation introduces minor errors and including the parabolic dispersion would slightly decrease the radiation in grazing angles, in particular, at very low temperatures.

\section{Analytical Model}
\label{sec:model}
The essence of the peculiar emission properties of TMDs is the presence of particle- and light-like excitons with transition matrix elements perpendicular and parallel to their motion, respectively. We now show that these features are captured by a simple analytical model. We include both exchange and SO coupling in the analysis. Furthermore, we show that a variational ansatz for the exciton allows for a quantitative estimate of the slope in linear light-like bands. To this end, we exploit the fact that states localized near the $K$- or $K'$-valleys are responsible for the $A$ and $B$ resonances. Eight such states exist due to the presence of two valleys as well as two spins for both valence and conduction bands, i.e.  $\sigma_v,\sigma_c \in \{\uparrow, \downarrow\}$. They can be grouped into spin-preserving, optically bright singlets $\ket{K_{\uparrow \uparrow}},\ket{K_{\downarrow \downarrow}},\ket{K'_{\uparrow \uparrow}},\ket{K'_{\downarrow \downarrow}}$ and spin-flipping, dark triplets $\ket{K_{\downarrow \uparrow}},\ket{K_{\uparrow \downarrow}},\ket{K'_{\downarrow \uparrow}},\ket{K'_{\uparrow \downarrow}}$. Here, $K_{\sigma_v \sigma_c}$ is a Wannier-like exciton localized to the $K$-valley and similarly for the $K'$-valley. The triplets can be omitted from the model since there is no SO-hybridisation in a two-band model with states of equal parity and, thus, no coupling to the bright singlets. At $Q=0$, the singlet $A$ and $B$ excitons have energies $E_A$ and $E_B$, respectively, with $E_B-E_A=2\Delta_{SO}$ the SO splitting. Without exchange effects, the center of mass motion simply adds a kinetic energy $\frac{\hbar^2 Q^2}{2 M}$, where $M$ is the sum of the hole and electron effective masses. In addition to SO coupling, we now include the exchange interaction leading to coupling between isolated exciton states. The intra- and intervalley exchange matrix elements are $V_x = V_x^{K K} = V_x^{K' K'}$ and $v_x = V_x^{K K'} = V_x^{K' K *}$, respectively. Furthermore, by choosing the center of mass momentum along $x$ so that $\vec{Q} = Q \unvec{x}$ we ensure that $v_x$ is real, as shown in App. \ref{sec:exchange}. Here, identities under interchange of valleys follow from time reversal symmetry. In this manner, the $4 \times 4$ Hamiltonian in the singlet basis including exchange coupling is $E_0(Q)\vec{I}+ \vec{H}$ with $E_0(Q) = \frac{1}{2}(E_A+E_B)+\frac{\hbar^2 Q^2}{2M}$ and
\begin{equation}
    \vec{H} = 
    \begin{bmatrix}
   -\Delta_{SO} & V_x & v_x& v_x \\
    V_x &  \Delta_{SO} & v_x & v_x \\
    v_x & v_x &   \Delta_{SO} & V_x \\
    v_x & v_x& V_x &  -\Delta_{SO}
    \end{bmatrix}.
    \label{eq:fullEigenboi}
\end{equation}
In App. \ref{sec:exchange}, we show that the intervalley exchange interaction is proportional to the center of mass momentum $v_x \propto Q$ for low values of $Q$. In principle, an intervalley Coulomb coupling $v_C$ also exists between the two valleys. However, as demonstrated in App. \ref{sec:exchange}, $v_C \propto Q^2$ for small values of $Q$, and we shall approximate it as $v_C = 0$. The energies found by diagonalising Eq.\eqref{eq:fullEigenboi} are then
\begin{equation}
\begin{split}
    E_{A_{p}}(Q) &= E_0(Q) + V_x + v_x - \Delta_{SO} \sqrt{1+(2\beta)^2}, \\
    E_{A_{l}}(Q) &= E_0(Q) + V_x - v_x - \Delta_{SO} \sqrt{1+(2\alpha)^2}, \\
    E_{B_{p}}(Q) &= E_0(Q) + V_x + v_x + \Delta_{SO} \sqrt{1+(2\beta)^2}, \\
    E_{B_{l}}(Q) &= E_0(Q) + V_x - v_x + \Delta_{SO} \sqrt{1+(2\alpha)^2},
\end{split}
\end{equation}
where $\alpha = \frac{V_x - v_x}{2 \Delta_{SO}}$ and $\beta = \frac{V_x + v_x}{2 \Delta_{SO}}$. The signs in front of $\Delta_{SO}$ and $v_x$ identify eigenvalues as belonging to the $A/B$-peaks and $p/l$-bands as indicated by the subscripts. In the $Q = 0$ limit, we see that $E_{A_{p}}(0) = E_{A_{l}}(0) = E_A$ and $E_{B_{p}}(0) = E_{B_{l}}(0) =E_B$. In addition, as shown in App. \ref{sec:exchange}, for small $Q$ we can write $v_x = - v_0 Q$ and $V_x = V_0 + v_0 Q$, where $v_0$ and $V_0$ are positive real-valued constants. Finally, in the case of large $\Delta_{SO}$, the exciton dispersions can be approximated as
\begin{equation}
\begin{split}
    E_{A_{p}}(Q) &= E_A + \frac{\hbar^2 Q^2}{2 M}, \\
    E_{A_{l}}(Q) &= E_A + \hbar Q V_- + \frac{\hbar^2 Q^2}{2 M_{-}}, \\
    E_{B_{p}}(Q) &= E_B + \frac{\hbar^2 Q^2}{2 M}, \\
    E_{B_{l}}(Q) &= E_B + \hbar Q V_+ + \frac{\hbar^2 Q^2}{2 M_{+}},
\end{split}
\end{equation}
where we introduce the effective velocities $V_\pm = 2 \frac{v_0}{\hbar} \left( 1 \pm \frac{V_0}{\Delta_{SO}} \right)$ and masses $M_{\pm} = M \left( 1 \pm \frac{4 v_0^2 M}{\hbar^2 \Delta_{SO}} \right)^{-1}$. This demonstrates that particle-like bands are parabolic while light-like bands contain a linear term. Note, also, that both effective masses and velocities of the light-like bands are affected by SO coupling such that the $A$ and $B$ light-like excitons have distinct band slopes and curvatures. Again, this finding agrees with our first-principles results.

\begin{figure}[ht]
    \centering
    \includegraphics[width=\linewidth]{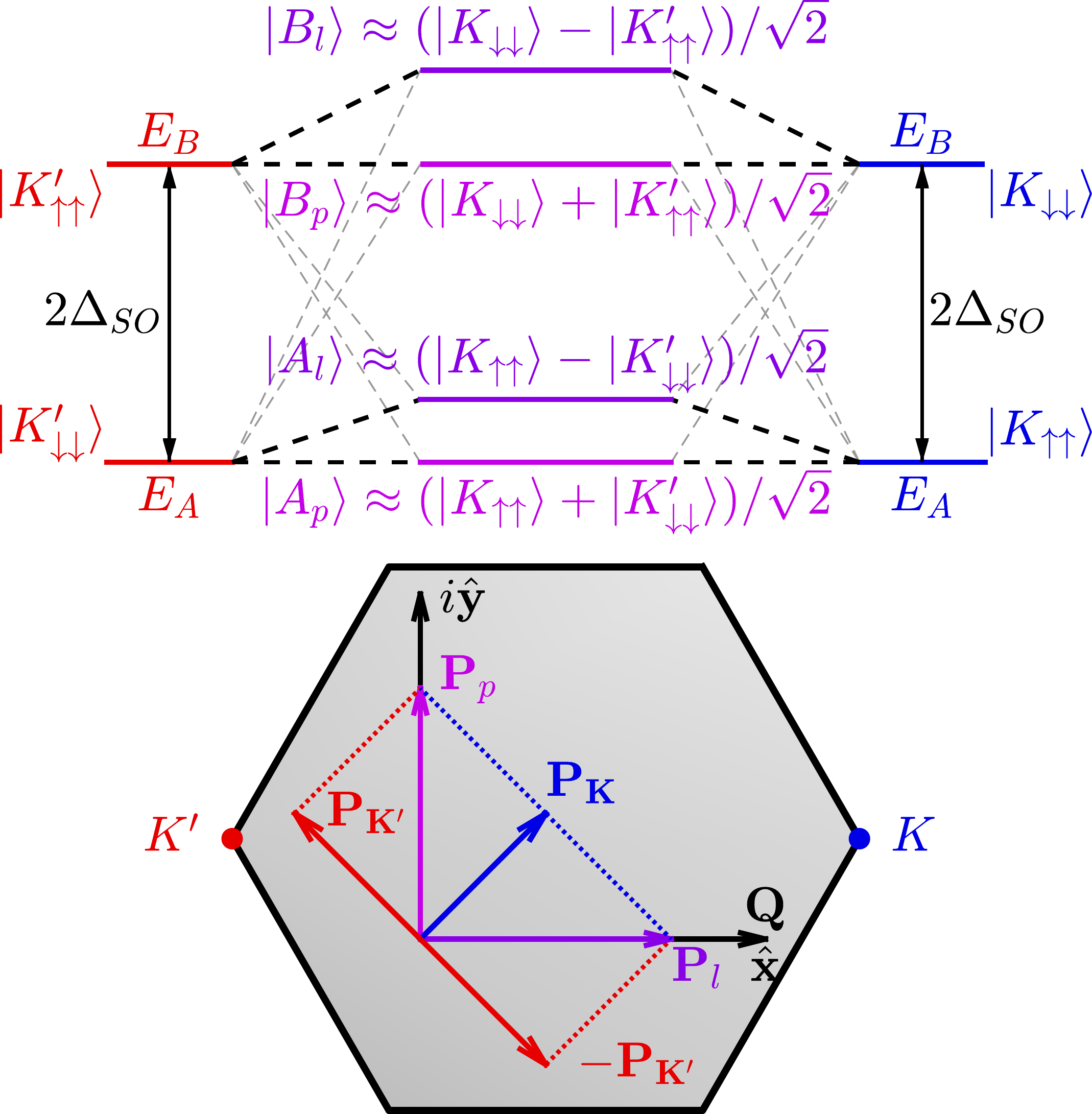}
    \caption{Schematic of the hybridisation between $K$ (blue) and $K'$ (red) excitons as a result of exchange for a center of mass momentum $\Vec{Q} = Q \unvec{x}$. Thick and thin dashed lines indicate large and small contributions, respectively, to the light-like (purple) and particle-like (pink) hybrids. The lower part shows exciton transition matrix elements for the light- and particle-like states oriented parallel and perpendicular to the center of mass momentum $\vec{Q}$, respectively.}
    \label{fig:momentumSum}
\end{figure}

The exact eigenvectors of Eq. \eqref{eq:fullEigenboi} are given in App. \ref{sec:fullSolve} but, in the limit $\Delta_{SO} \gg V_x + \abs{v_x}$, they can be approximated as
%\begin{equation}
%\begin{split}
%    \Ket{A_{p}} &= \frac{\ket{K_{\uparrow \uparrow}} - \beta \ket{K_{\downarrow \downarrow}} + \ket{K'_{\downarrow \downarrow}} - \beta \ket{K'_{\uparrow \uparrow}}}{\sqrt{2 + 2\beta^2}}, \\
%    \Ket{A_{l}} &= \frac{\ket{K_{\uparrow \uparrow}} - \alpha \ket{K_{\downarrow \downarrow}} - \ket{K'_{\downarrow \downarrow}} + \alpha \ket{K'_{\uparrow \uparrow}}}{\sqrt{2 + 2\alpha^2}}, \\
%    \Ket{B_{p}} &= \frac{\ket{K_{\downarrow \downarrow}} + \beta \ket{K_{\uparrow \uparrow}} + \ket{K'_{\uparrow \uparrow}} + \beta \ket{K'_{\downarrow \downarrow}}}{\sqrt{2 + 2\beta^2}}, \\
%    \Ket{B_{l}} &= \frac{\ket{K_{\downarrow \downarrow}} + \alpha \ket{K_{\uparrow \uparrow}} - \ket{K'_{\uparrow \uparrow}} - \alpha \ket{K'_{\downarrow \downarrow}}}{\sqrt{2 + 2\alpha^2}}.
%\end{split}
%\end{equation}
\begin{equation}
\begin{split}
    \Ket{A_{p}} &= 2^{-1/2} \left( \ket{K_{\uparrow \uparrow}} - \beta \ket{K_{\downarrow \downarrow}} + \ket{K'_{\downarrow \downarrow}} - \beta \ket{K'_{\uparrow \uparrow}} \right), \\
    \Ket{A_{l}} &= 2^{-1/2} \left( \ket{K_{\uparrow \uparrow}} - \alpha \ket{K_{\downarrow \downarrow}} - \ket{K'_{\downarrow \downarrow}} + \alpha \ket{K'_{\uparrow \uparrow}} \right), \\
    \Ket{B_{p}} &= 2^{-1/2} \left( \ket{K_{\downarrow \downarrow}} + \beta \ket{K_{\uparrow \uparrow}} + \ket{K'_{\uparrow \uparrow}} + \beta \ket{K'_{\downarrow \downarrow}} \right), \\
    \Ket{B_{l}} &= 2^{-1/2} \left( \ket{K_{\downarrow \downarrow}} + \alpha \ket{K_{\uparrow \uparrow}} - \ket{K'_{\uparrow \uparrow}} - \alpha \ket{K'_{\downarrow \downarrow}} \right).
\end{split}
\end{equation}
The SO splitting $2 \Delta_{SO}$ is substantial ($\approx 0.6 \, \text{eV}$) in tungsten-based TMDs and significantly smaller ($\approx 0.2 \, \text{eV}$) in molybdenum compounds \cite{article:C2DB}. Thus, in the former case, $\abs\alpha,\abs\beta\ll 1$ and eigenstates are approximately $\ket{K_{\uparrow \uparrow}}\pm\ket{K'_{\downarrow \downarrow}}$ or $\ket{K_{\downarrow \downarrow}}\pm\ket{K'_{\uparrow \uparrow}}$, all of which yield bright excitons, c.f.  Fig. \ref{fig:momentumSum}. On the other hand, for the molybdenum based TMDs, where $\Delta_{SO}$ is significantly smaller, the $A$-peak eigenstates will tend towards a triplet state, making them less bright. Thus, the exchange term implies that molybdenum based TMDs show a more pronounced hybridisation between the two spin-conserving transitions at the $K$-symmetry points. This is important since the exchange term is particularly affected by external screening. It follows that hybridisation will be reduced in a dielectric environment impacting, in turn, significantly excitonic states and transition matrix elements.

The appearance of strictly parallel- and perpendicular exciton transition matrix elements can now be demonstrated in the limit of small $Q$, where the time reversal symmetry is upheld. The transition matrix elements for the band transitions of TMDs at the $K$ and $K'$ high symmetry points are $\vec{p}_{\Vec{K}}=p_0 \unvec{e}_-$ and $\vec{p}_{\vec{K'}}=-p_0 \unvec{e}_+$ as derived by Taghizadeh \textit{et al.} \cite{article:selectionRules}. Here, $\unvec{e}_{\pm}\equiv(\unvec{x}\pm i\unvec{y})/\sqrt{2}$ are the chiral unit vectors introduced earlier, and the sign follows from time reversal symmetry. For $\vec{k}$ in the neighborhood of $\vec{K}$, we write $\Vec{p}_{\vec{k}}\approx p_{\vec{k}}\unvec{e}_-$. The corresponding exciton matrix element for the Wannier-like excitons in the $K$-valley is then approximated as
\begin{equation}
\begin{split}
    \vec{P}_{\Vec{K}}(\Vec{Q})\approx \frac{A}{(2 \pi)^2}\int \psi_0(\Vec{k} - \vec{K})\Vec{p}_{\vec{k}} d^2 \vec{k}\approx P_0 \unvec{e}_-,
    \label{eq:momentumRequirement}
\end{split}
\end{equation}
where $\psi_0(\Vec{k} - \Vec{K})$ is the Wannier wave function for an isolated exciton in the $K$ valley with center of mass momentum $Q = 0$. For the other valley, $\vec{P}_{\Vec{K'}}(\Vec{Q})\approx -P_0 \unvec{e}_+$. Now, by approximating the transition matrix elements for the Wannier-like excitons of different spin, we find $\vec{P}_{K_{\uparrow \uparrow}} = \vec{P}_{K_{\downarrow \downarrow}}\approx P_0 \unvec{e}_-$ and $\vec{P}_{K'_{\uparrow \uparrow}} = \vec{P}_{K'_{\downarrow \downarrow}}\approx -P_0 \unvec{e}_+$. Thus, combining the eigenvectors with the transition matrix elements yields the exciton transition matrix elements
%\begin{equation}
%\begin{split}
%    \Vec{P}_{A_p} &= \frac{1 - \beta }{\sqrt{2 + 2\beta^2}} \left(\vec{P}_{K0} -\vec{P}_{K0}^* \right), \\
%    \Vec{P}_{A_l} &= \frac{1 - \alpha}{\sqrt{2 + 2\alpha^2}} \left(\vec{P}_{K0} + \vec{P}_{K0}^* \right), \\
%    \Vec{P}_{B_p} &= \frac{1 + \beta }{\sqrt{2 + 2\beta^2}} \left(\vec{P}_{K0} - \vec{P}_{K0}^* \right), \\
%    \Vec{P}_{B_l} &= \frac{1 + \alpha}{\sqrt{2 + 2\alpha^2}} \left(\vec{P}_{K0} + \vec{P}_{K0}^* \right),
%\end{split}
%\end{equation}
%where Eq. \eqref{eq:momentumRequirement} has been used to write the transition matrix elements in the $K'$-valley in terms of $K$-valley properties. Using the phase requirement of Eq. \eqref{eq:phaseDefinitions} as well as Eq. \eqref{eq:alireza} then reduces the exciton transition matrix elements to
\begin{equation}
\begin{split}
    \Vec{P}_{A_p} &= -i(1 - \beta)  P_0 \unvec{y}, \\
    \Vec{P}_{A_l} &= (1 - \alpha)  P_0 \unvec{x}, \\
    \Vec{P}_{B_p} &= -i(1 + \beta) P_0 \unvec{y}, \\
    \Vec{P}_{B_l} &= (1 + \alpha) P_0 \unvec{x}.
    \label{eq:momentumMatrixElements}
\end{split}
\end{equation}
That is, the exciton transition matrix element of the light-like eigenstates will be parallel to $\Vec{Q} = Q \unvec{x}$, while for the particle-like eigenstates it will be perpendicular. This is illustrated in Fig. \ref{fig:momentumSum}, where the exciton wave functions are simple symmetric and anti-symmetric superposition of the $K$ and $K'$ states. We stress that, under the assumption of circular symmetry, this is general for any direction of $\Vec{Q}$. Furthermore, the exciton transition matrix elements will be reduced as $V_x + v_x$ approaches $\Delta_{SO}$ for the $A$-peak excitons, while they increase slightly for the $B$-peak excitons, due to the different signs in front of $\alpha$ and $\beta$. In turn, this means that substrate screening can have a significant effect on the transition matrix elements. Our numerical calculations of the absolute squared transition matrix element $||\Vec{P}||^2$ for TMDs on a quartz substrate show a $30\%$ increase for {\MoS} and {\MoSe}, while they show a mere $2\%$ increase for the two tungsten based TMDs. %This effect can be seen in Fig. \ref{fig:2D_dispersion}c, where the dispersions for {\MoS} in two common environments (SiO$_2$ substrate and hBN encapsulation) is shown.

\begin{figure}[ht]
    \centering
    \includegraphics[trim={0.0cm 3.0cm 2.0cm 4.1cm},clip,width=\linewidth]{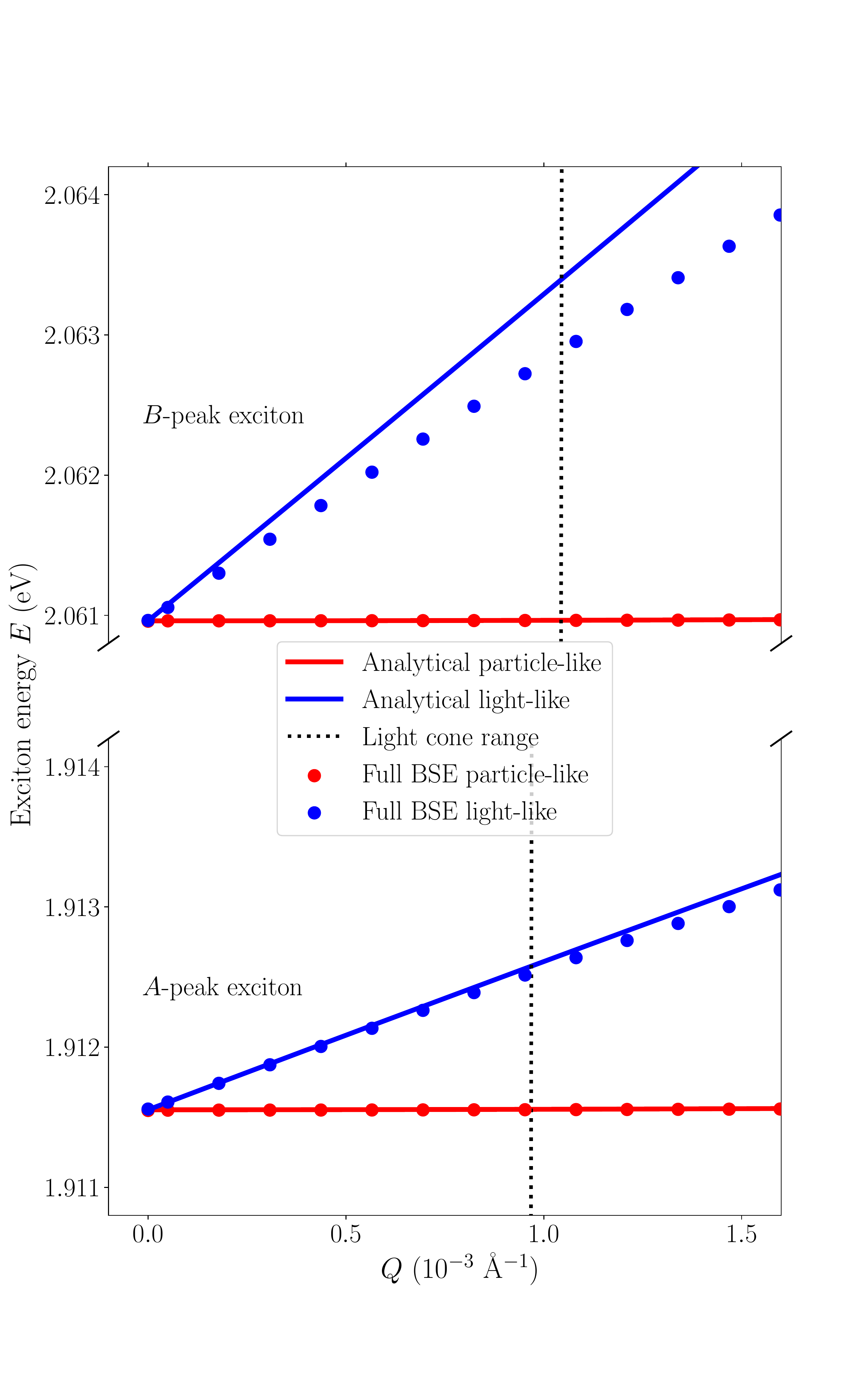}
    \caption{Comparison of exciton dispersions in {\MoS} in the analytical model (solid lines) and the full BSE solution (dots). Both light- and particle-like bands are shown and the range of the light cone $Q=\omega/c$ is indicated by vertical dotted lines.}
    \label{fig:effectiveComp}
\end{figure}

To compare this analytical model to the \textit{ab initio} model, we shall apply a simple estimate using the variational ansatz for the Wannier-like exciton in position space
\begin{equation}
     \psi_0(\Vec{r}) =N(\exp{-a r} - b \exp{-\gamma a r}),
\end{equation}
which has proven effective at describing the ground exciton state in {\MoS} \cite{article:starkShift}. The parameter $N$ is a normalisation constant, and the parameters $a$, $b$ and $\gamma$ are found by minimising the energy in a Wannier model including the Keldysh potential. In Fourier space, this wave function is 
\begin{equation}
    \psi_0(\Vec{k}) = N \left(\frac{a}{(a^2+k^2)^{3/2}} - \frac{b \gamma a}{(\gamma^2 a^2+k^2)^{3/2}} \right).
\end{equation}

%Furthermore, we approximate $\epsilon_{cv\Vec{k}} \approx \sqrt{E_g^2 + (2 \hbar v_F k)^2}$ and $||\Vec{p}_K|| \approx m_e v_F \sqrt{1 + E_g^2/\epsilon_{cv\Vec{k}}^2}$, where $2 v_F = \sqrt{\frac{E_g}{\mu}}$. Using this the integral of Eq. \eqref{eq:alphaint} can be solved, yielding 
%\begin{equation}
%\begin{split}
%    v_x &= -|F(\hbar \kappa / (2 \mu v_F)|^2 \frac{Q^2 w(Q)}{4 \pi} \exp{2i\phi}, \\
%    \text{where:} \quad F(x) &= \frac{x (\sqrt{2} - x)}{1 - x^2} - \frac{x^2}{(1-x^2)^{3/2}} \\
%    &\times \arctan{\left( \frac{(\sqrt{2} - x)\sqrt{1-x^2})}{x+\sqrt{2}(1-x^2)} \right)},
%\end{split}
%\end{equation}

%\begin{equation}
%\begin{split}
%    v_x &= -\left| \frac{F(\hbar \kappa / (2 \mu v_F)}{\kappa} + c \frac{F(\hbar \gamma \kappa / (2 \mu v_F)}{\gamma \kappa} %\right|^2 \frac{Q^2 w(Q)}{8N^2} \exp{2i\phi}, \\
%\end{split}
%\end{equation}
%where
%\begin{equation}
%    \begin{split}
%        \quad F(x) &= \frac{x (\sqrt{2} - x)}{1 - x^2} - \frac{x^2}{(1-x^2)^{3/2}} \\
%    &\times \arctan{\left( \frac{(\sqrt{2} - x)\sqrt{1-x^2})}{x+\sqrt{2}(1-x^2)} \right)},
%    \end{split}
%\end{equation}
%and $w(Q) = \frac{e^2}{2 \eps_0 Q}$. Thus, applying values for {\MoS} $E_g = 2.53\,\text{eV}$, $\mu = 0.237 m_e$, $r_0 = 39.5\,\text{Å}$, we obtain $|v_x| = 0.845 \, \text{eV} \, \text{Å} \, Q$. 
To calculate the value of $v_{0}$, we need the transition matrix elements and energies for the band transitions, which can obtained from the DFT calculation. Applying these in equation \eqref{eq:alphaint}, gives a value of $v_{0} = 0.845 \, \text{eV} \, \text{Å}$. To calculate the value of $V_{0}$, we need to include all the $\Vec{G}$-vectors of Eq. \eqref{eq:exchangePart}, and average it over our variational ansatz $\psi_0$, that is 
\begin{equation}
    V_{0} = \int \int \psi_0^*(\vec{k}) v_x(\Vec{k},\Vec{k}',Q=0) \psi_0(\vec{k}') d^2 \Vec{k} d^2 \Vec{k}'.
\end{equation}
Applying our DFT calculation again, we obtain a value of $V_{0} = 27.4 \, \text{meV}$. Using these values to calculate the dispersion for {\MoS} then yields the plot of Fig. \ref{fig:effectiveComp}, which shows good agreement between the two approaches in the limit of small perturbation $Q$ for the $A$-peak, while deviating slightly for the $B$-peak. This deviation is, however, quite small compared to the approximations made throughout this section, such as the Wannier-Mott approximation, the variational ansatz, and the limited amount of included bands and exciton states. In fact, a quantitative description of exciton band structures and transition matrix elements can be obtained in this manner, without actually solving the BSE.

\section{Summary}
%In summary, we  first-principles calculations of the exciton band structure for monolayer {\MoS}, {\MoSe}, {\WS}, and {\WSe}. From this, we find a light-like and particle-like exciton band representing the $K$-valley excitons, in agreement with previously established results. These excitons show little to no angular dependence for thermally allowed values $\Vec{Q}$. Furthermore, we find that the transition matrix element of each state is either purely perpendicular or parallel to the center of mass momentum, for the particle- and light-like exciton, respectively. Based on the obtained dispersions, we calculate the radiative lifetimes of the monolayer TMDs including both the particle-like and light-like bands, and in doing so find good agreement with available experimental results when taking into account dielectric screening and the Purcell effect. To this purpose, we perform calculations of the radiative lifetimes in a dielectric medium illustrating the significance of the dielectric screening on the transition matrix elements and of the Purcell effect on the photon phase space. Finally, we derive an effective model capable of capturing the essential physics related to the light-like and particle-like excitons, the $\Vec{Q}$-dependency of their transition matrix elements, as well as the exchange couplings interplay with spin orbit coupling for the brightness of an exciton.
In summary, we have studied excitons and their optical emission in the monolayer TMDs {\MoS}, {\MoSe}, {\WS}, and {\WSe} using a first-principles DFT+BSE approach. Our emphasis is on excitons moving with a finite center of mass momentum and their effect on emission. In TMDs, exciton energies form characteristic particle- and light-like bands. The latter has previously been ignored but we demonstrate that clear experimental fingerprints are expected in both temperature and angular dependence of the emission. In particular, optical selection rules relating motion and emission directions are found to be essentially opposite for particle- and light-like excitons. Signatures of light-like excitons appear due the different thermal populations of the exciton bands combined with their selection rules. Moreover, we describe the effects of screening by a dielectric environment on both exciton band structure and optical emission. Our results are in good agreement with available experimental radiative lifetimes. Finally, a simple analytical model including exchange and spin-orbit coupling of valley excitons is shown to capture the essential physics.

\begin{acknowledgments}
M.O.S. and T.G.P. are supported by the CNG center under the Danish National Research Foundation, project DNRF103. 
\end{acknowledgments}

\onecolumngrid

\appendix
\section{Intervalley Exchange and Coulomb Interaction} 
\label{sec:exchange}
To find an expression for the $K-K'$ exciton exchange interaction, we start from Bloch's theorem to write the lattice-periodic part of the wave function $u_{n,\Vec{k}}(\Vec{r}) = \exp{-i \Vec{k} \cdot \Vec{r}} \psi_{n,\Vec{k}}(\Vec{r})$, which is then expanded in 2D Fourier coefficients using the periodicity of the unit cell
\begin{equation}
   u_{n,\Vec{k}}(\Vec{r}) = \frac{1}{\sqrt{A}} \sum_{\Vec{G}} C_{n,\Vec{k}}(\Vec{G}, z) \exp{i \Vec{G} \cdot \vec{\rho}},
\end{equation}
where $\Vec{\rho}$ is a 2D spatial coordinate and $A$ the TMD area. Next, products of these are Fourier decomposed, such that
\begin{equation}
u_{n,\Vec{k}}^*(\Vec{r}) u_{m,\Vec{k}'}(\Vec{r}) = \frac{1}{A} \sum_{\Vec{G}} i_{n\Vec{k},m\Vec{k}'}(\Vec{G}, z) \exp{i \Vec{G} \cdot \vec{\rho}},
\end{equation}
where the Fourier coefficients are given as 
\begin{equation}
    i_{n\Vec{k},m\Vec{k}'}(\Vec{G}, z) = \int_{A} u_{n,\Vec{k}}^* \exp{-i \Vec{G} \cdot \vec{\rho}} u_{m,\Vec{k}'} d^2 \vec{\rho} = \sum_{\Vec{G}'} C_{n \Vec{k}}^* (\Vec{G}' - \Vec{G}, z) C_{m \Vec{k}'} (\Vec{G}', z).
\end{equation}
Likewise, the Coulomb interaction $\op{W}(\Vec{r},\Vec{r}')$ is Fourier decomposed in the entire 2D-plane
\begin{equation}
    \op{W}(\Vec{r},\Vec{r}') = \frac{1}{A} \sum_{\Vec{q},\Vec{G}} w(\Vec{q}+\Vec{G}, z, z') \exp{i(\Vec{q}+\Vec{G})\cdot(\vec{\rho}-\vec{\rho}')},
\end{equation}
where $\Vec{q}$ is a 2D vector confined to the Brillouin zone, and $w(\Vec{q}+\Vec{G}, z, z')$ is the Coulomb potential in 2D Fourier space. Taking the 2D limit of the screening, in accordance with the Keldysh approximation, we get
\begin{equation}
    w(\Vec{q}+\Vec{G}, z, z') \approx w(\Vec{q}+\Vec{G}, 0, 0),
\end{equation}
where the TMD has been placed at $z=0$. The elements of the Coulomb interaction are
\begin{equation}
\begin{split}
    \braket{1,2|\op{W}(\Vec{r},\Vec{r}')|3,4} &= \frac{1}{A^3} \sum_{\Vec{q}, \Vec{G}, \vec{G}_{1,3},\vec{G}_{2,4}} w(\Vec{q}+\Vec{G}, 0, 0) \int_A \exp{i \left(\Vec{G}_W + \Vec{G}_{1,3} + \Vec{q} + \Vec{k}_3 - \Vec{k}_1 \right) \cdot \vec{\rho}} d^2 \vec{\rho} \int_{-\infty}^{\infty} i_{1,3}(\Vec{G}_{1,3}, z) d z \\  
    &\times \int_A \exp{-i \left(\Vec{G} - \Vec{G}_{2,4} + \Vec{q} - \Vec{k}_4 + \Vec{k}_2 \right) \cdot \vec{\rho}} d^2 \vec{\rho} \int_{-\infty}^{\infty} i_{2,4}(\Vec{G}_{2,4}, z) d z .
\end{split}
\end{equation}
Due to the integrals over the area, the non-zero terms in the sums will have vanishing exponents in the integrand. Furthermore, since $\Vec{q}$ and $\Vec{k}$ are restricted to the Brillouin zone, we get $\Vec{q} = \Vec{k}_4 - \Vec{k}_2 = \Vec{k}_1 - \Vec{k}_3$ and $\Vec{G} = -\Vec{G}_{1,3} = \Vec{G}_{2,4}$.
We now define $I_{1, 2}(\Vec{G}) \equiv \int_{-\infty}^{\infty} i_{1,2}(\Vec{G}, z) d z$, such that the Coulomb ($C$) and exchange ($x$) kernels become
\begin{align}
    v_C(\Vec{k},\Vec{k}',\vec{Q}) &= \braket{v\Vec{k}',c\Vec{k} + \Vec{Q}|\op{W}_C(\Vec{r},\Vec{r}')|v\Vec{k},c\Vec{k}' + \Vec{Q}} =  \frac{1}{A} \sum_{\Vec{G}} w_C(\Vec{k}' - \Vec{k} + \vec{G}) I_{v \Vec{k}, v \Vec{k}'}^*(\Vec{G}) I_{c \Vec{k} + \Vec{Q}, c \Vec{k}' + \Vec{Q}}(\Vec{G}), 
    \label{eq:coulombPart}\\
    v_x(\Vec{k},\Vec{k}',\vec{Q}) &= \braket{v\Vec{k},c\Vec{k}' + \Vec{Q}|\op{W}_x(\Vec{r},\Vec{r}')|c\Vec{k} + \Vec{Q}, v\Vec{k}'} =  \frac{1}{A} \sum_{\Vec{G}} w_x(\vec{G} - \Vec{Q}) I_{c \Vec{k} + \Vec{Q}, v \Vec{k}}^*(\Vec{G}) I_{c \Vec{k}' + \Vec{Q}, v \Vec{k}'}(\Vec{G}),
    \label{eq:exchangePart}
\end{align}
where Hermiticity $I_{1, 2}(\Vec{G}) = I_{2, 1}^*(-\Vec{G})$ has been applied. Furthermore, we expand Hamilton matrix elements to first order in $\Vec{Q}$, such that
\begin{equation}
    H_{\Vec{G},\Vec{G}'} (\Vec{k} + \Vec{Q}) \approx H_{\Vec{G},\Vec{G}'} (\Vec{k}) + \frac{\hbar^2}{m_e} \Vec{Q} \cdot (\Vec{k} + \Vec{G}) \delta_{\Vec{G},\Vec{G}'}.
\end{equation}
Applying the two-band approximation and using first-order perturbation theory for the perturbed conduction states at the $K$ and $K'$ points then yields the plane wave coefficients
\begin{equation}
    C_{c,\Vec{k}+\Vec{Q}}(\Vec{G}, z) \approx C_{c,\Vec{k}}(\Vec{G}, z) + \frac{\hbar}{m_e}\frac{\Vec{Q} \cdot \Vec{p}_{\vec{k}}}{\epsilon_{\vec{k}}} C_{v,\Vec{k}}(\Vec{G}, z),
    \label{eq:planeBois}
\end{equation}
where $\epsilon_{\vec{k}}=\epsilon_{c,\vec{k}}-\epsilon_{v,\vec{k}}$ is the band-to-band transition energy and the $z$-component of $\Vec{p}_{\vec{k}}$ is zero. As a consequence, the Coulomb kernel will vary approximately as $Q^2$, since the screening is independent of $\vec{Q}$. In the so called slow-rapid approximation, we neglect all terms except $\Vec{G} = 0$. This is a valid approximation for the intervalley exchange since only the $\Vec{G} = 0$ term will be proportional to $Q$ for small $Q$, while $\Vec{G} \neq 0$ terms only contribute to order $Q^2$ and higher as a constant $Q^0$ contribution would break time reversal symmetry. Using Eq. \eqref{eq:planeBois} and the orthonormality of wave functions, we get
\begin{equation}
    \label{eq:momentumEquality}
    I_{c \Vec{k} + \Vec{Q}, v \Vec{k}}(0)  \approx \frac{\hbar}{m_e}\frac{\Vec{Q} \cdot \Vec{p}_{\vec{k}}}{\epsilon_{\vec{k}}},
\end{equation}
and the exchange kernel in the slow-rapid approximation then becomes
\begin{equation}
    v_x(\Vec{k},\Vec{k}',\vec{Q}) = \frac{1}{A} w_x(Q) I_{c \Vec{k} + \Vec{Q}, v \Vec{k}}^*(0) I_{c \Vec{k}' + \Vec{Q}, v \Vec{k}'}(0).
\end{equation}
The $K-K'$ exciton exchange interaction $v_x$ is, thus, given by
\begin{equation}
\begin{split}
        v_x &= \frac{w_x(Q)}{(2 \pi)^2} \int I^*_{c \Vec{k} + \Vec{Q}, v \Vec{k}}(0) \psi_0(\Vec{k}-\Vec{K}) d^2 \Vec{k} \int I_{c \Vec{k} + \Vec{Q}, v \Vec{k}}(0) \psi_0(\Vec{k}-\Vec{K}') d^2 \Vec{k} \\
        &\approx \frac{w_x(Q)}{(2 \pi)^2} \frac{\hbar^2}{m_e^2} \int \frac{\Vec{Q} \cdot \Vec{p}_{\vec{k}}^*}{\epsilon_{\vec{k}}} \psi_0(\Vec{k}-\Vec{K}) d^2 \Vec{k} \int \frac{\Vec{Q} \cdot \Vec{p}_{\vec{k}}}{\epsilon_{\vec{k}}} \psi_0(\Vec{k}-\Vec{K}') d^2 \Vec{k},
\end{split}
\end{equation}
where Eq. \eqref{eq:momentumEquality} has been applied. Here, $\psi_0(\Vec{k}-\Vec{K})$ is the unperturbed Wannier wave function centered at $\Vec{K}$. Applying the center of mass momentum $\Vec{Q} = Q \unvec{x}$ and transition matrix elements $\vec{p}_{\Vec{k}} \approx p_{\vec{k}} \unvec{e}_-$ in the vicinity of $K$ and $\vec{p}_{\Vec{k}} \approx -p_{\vec{k}} \unvec{e}_+$ in the vicinity of $K'$, used throughout Sec. \ref{sec:model}, we get
%\begin{equation}
%    v_x \approx \frac{w_x(Q)}{(2 \pi)^2} \left| \frac{\hbar}{m_e} \int \frac{p_{\vec{k}}}{\epsilon_{\vec{k}}} \psi_0(\Vec{k} - \Vec{K}) d^2 \Vec{k} \right|^2 (\vec{Q} %\cdot\unvec{p}_{\Vec{K}}^*) (\vec{Q} \cdot \unvec{p}_{\Vec{K'}}).
%    \label{eq:alphaint}
%\end{equation}
%Applying the center of mass momentum $\Vec{Q} = Q \unvec{x}$ and transition matrix element unit vectors $\unvec{p}_{\Vec{K}} = - \unvec{p}_{\Vec{K'}}^* = \unvec{e}_-$, used throughout Sec. \ref{sec:model}, we get
\begin{equation}
    v_x \approx -\frac{w_x(Q)}{2 (2 \pi)^2} \left| \frac{\hbar Q}{m_e} \int \frac{p_{\vec{k}}}{\epsilon_{\vec{k}}} \psi_0(\Vec{k} - \Vec{K}) d^2 \Vec{k} \right|^2.
    \label{eq:alphaint}
\end{equation}
Note that $p_{\vec{k}}$ is real and any complex phase is handled by $\unvec{e}_\pm$. It follows that $v_x$ is real-valued, due to the choice $\Vec{Q} = Q \unvec{x}$ as allowed under full rotational symmetry. Choosing a center of mass momentum with a different orientation relative to the coordinate system would make $v_x$ complex. The phase, however, would be precisely cancelled by an equal and opposite phase of the transition matrix element.
\section{Arbitrary Spin-Orbit Coupling}
\label{sec:fullSolve}
In the main text, the limit of dominant SO coupling is discussed for simplicity. In fact, the general case can be handled analytically. The full Hamiltonian for the $K$-$K'$ interaction is given in Eq. \eqref{eq:fullEigenboi}. Solving the associated eigenvalue problem yields the eigenvalues
\begin{equation}
\begin{split}
    E_{A_{p}}(Q) &= E_0(Q) + V_x + v_x - \Delta_{SO} \sqrt{1+(2\beta)^2}, \quad
    E_{A_{l}}(Q) = E_0(Q) + V_x - v_x - \Delta_{SO} \sqrt{1+(2\alpha)^2}, \\
    E_{B_{p}}(Q) &= E_0(Q) + V_x + v_x + \Delta_{SO} \sqrt{1+(2\beta)^2}, \quad
    E_{B_{l}}(Q) = E_0(Q) + V_x - v_x + \Delta_{SO} \sqrt{1+(2\alpha)^2},
\end{split}
\end{equation}
where $\alpha = \frac{V_x - v_x}{2 \Delta_{SO}}$ and $\beta = \frac{V_x + v_x}{2 \Delta_{SO}}$ as in Sec. \ref{sec:model}. The corresponding eigenvectors are
\begin{equation}
\begin{split}
    \Ket{A_{p}} &= \frac{\big(1+\sqrt{1+(2\beta)^2} \big) \ket{K_{\uparrow \uparrow}} - 2 \beta \ket{K_{\downarrow \downarrow}} + \big(1+\sqrt{1+(2\beta)^2} \big) \ket{K'_{\downarrow \downarrow}} - 2\beta \ket{K'_{\uparrow \uparrow}} }{\sqrt{2 \big(1+\sqrt{1+(2\beta)^2} \big)^2+8 \beta^2}}, \\
    \Ket{A_{l}} &= \frac{\big(1+\sqrt{1+(2\alpha)^2} \big) \ket{K_{\uparrow \uparrow}} - 2 \alpha \ket{K_{\downarrow \downarrow}} - \big(1+\sqrt{1+(2\alpha)^2} \big) \ket{K'_{\downarrow \downarrow}} + 2 \alpha \ket{K'_{\uparrow \uparrow}}}{\sqrt{2 \big(1+\sqrt{1+(2\alpha)^2} \big)^2+8\alpha^2}}, \\
    \Ket{B_{p}} &= \frac{\big(1-\sqrt{1+(2\beta)^2} \big) \ket{K_{\uparrow \uparrow}} - 2 \beta \ket{K_{\downarrow \downarrow}} + \big(1-\sqrt{1+(2\beta)^2} \big) \ket{K'_{\downarrow \downarrow}} - 2 \beta \ket{K'_{\uparrow \uparrow}} }{\sqrt{2 \big(1-\sqrt{1+(2\beta)^2} \big)^2+8 \beta^2}}, \\
    \Ket{B_{l}} &= \frac{\big(1-\sqrt{1+(2\alpha)^2} \big) \ket{K_{\uparrow \uparrow}} - 2\alpha \ket{K_{\downarrow \downarrow}} - \big(1-\sqrt{1+(2\alpha)^2} \big) \ket{K'_{\downarrow \downarrow}} + 2 \alpha \ket{K'_{\uparrow \uparrow}}}{\sqrt{2 \big(1-\sqrt{1+(2\alpha)^2} \big)^2+8 \alpha^2}}.
\end{split}
\end{equation}
It should be noted that, in the limit of $\Delta_{SO} \rightarrow 0$, we get the usual superpositions of the $K$ and $K'$ singlet- and triplet states. To show that the exciton transition matrix elements are parallel and perpendicular to $\Vec{Q}$, we use $\vec{P}_{K_{\uparrow \uparrow}} = \vec{P}_{K_{\downarrow \downarrow}}\approx P_0\unvec{e}_-$ and $\vec{P}_{K'_{\uparrow \uparrow}} = \vec{P}_{K'_{\downarrow \downarrow}}\approx -P_0\unvec{e}_+$ similarly to Sec. \ref{sec:model}.
\begin{equation}
\begin{split}
    \Vec{P}_{A_p} &= -i P_0 \frac{1+\sqrt{1+(2\beta)^2} - 2 \beta }{\sqrt{\big(1+\sqrt{1+(2\beta)^2} \big) ^2+4 \beta^2}} \unvec{y}, \\
    \Vec{P}_{A_l} &= P_0 \frac{1+\sqrt{1+(2\alpha)^2} - 2 \alpha }{\sqrt{\big(1+\sqrt{1+(2\alpha)^2} \big)^2+4 \alpha^2}} \unvec{x}, \\
    \Vec{P}_{B_p} &= -i P_0 \frac{1-\sqrt{1+(2\beta)^2} - 2 \beta }{\sqrt{\big(1-\sqrt{1+(2\beta)^2} \big)^2+4 \beta^2}} \unvec{y}, \\
    \Vec{P}_{B_l} &= P_0 \frac{1-\sqrt{1+(2\alpha)^2} - 2 \alpha }{\sqrt{\big(1-\sqrt{1+(2\alpha)^2} \big)^2+4 \alpha^2}} \unvec{x}.
\end{split}
\end{equation}
That is, the transition matrix elements are either parallel or perpendicular to the center of mass momentum $\Vec{Q} = Q \unvec{x}$

\twocolumngrid

\bibliography{bibliography} % Produces the bibliography via BibTeX.

\end{document}